\documentclass[10pt]{IEEEtran}
\usepackage{cite}
\usepackage{amsmath,amssymb,amsfonts}
\usepackage{algorithmic}
\usepackage{graphicx}
\usepackage{svg}     
\usepackage{textcomp}
\usepackage{xcolor}
\usepackage{booktabs}
\usepackage{quantikz}
\usepackage{adjustbox}
\usepackage{float}
\usepackage[font=footnotesize,skip=0pt]{caption}
\usepackage{subcaption}
\usepackage{comment}
\usepackage{url}
\usepackage{xurl}
\usepackage{hyperref}
\usepackage{breakurl}

\usepackage{enumitem}
\setlist[itemize]{leftmargin=*}%
\setlist[enumerate]{leftmargin=*}%

\definecolor{layerblue}{RGB}{100, 180, 255}
\definecolor{embeddinggreen}{RGB}{213, 233, 212}
\definecolor{sqrtx}{RGB}{215, 232, 254}

\definecolor{sc1bg}{RGB}{213, 233, 212}
\definecolor{sc2bg}{RGB}{215, 232, 254}
\definecolor{sc3bg}{RGB}{255, 230, 204}
\definecolor{sc4bg}{RGB}{250, 230, 255}
\definecolor{sc5bg}{RGB}{248, 206, 204}
\definecolor{embedcolor}{RGB}{80, 200, 120}
\definecolor{rxcolor}{RGB}{255, 100, 100}
\definecolor{rycolor}{RGB}{100, 180, 255}
\definecolor{rzcolor}{RGB}{200, 150, 255}
\definecolor{cnotcolor}{RGB}{255, 200, 80}
\definecolor{measurecolor}{RGB}{180, 180, 180}
\definecolor{cutred}{RGB}{220, 0, 0}
\newcommand{\cutsym}{\textcolor{cutred}{\scalebox{0.9}{\boldmath$\times$}}}

\tikzset{
    rzgate/.style={fill=embeddinggreen,draw=black!80,thick,rounded corners=2pt,minimum width=1.2cm,minimum height=0.7cm,text width=1.0cm,align=center,font=\small},
    sqrtxgate/.style={fill=sqrtx,draw=black,thick,rounded corners=2pt,minimum width=1.0cm,minimum height=0.7cm},
    metergate/.style={fill=measurecolor,draw=black,thick,rounded corners=2pt,minimum height=0.7cm}
}

\allowdisplaybreaks
\def\BibTeX{{\rm B\kern-.05em{\black i\kern-.025em b}\kern-.08em
    T\kern-.1667em\lower.7ex\hbox{E}\kern-.125emX}}

\usepackage[compact]{titlesec}
\usepackage{stfloats}

\titlespacing\section{0pt}{0.2\baselineskip}{0.12\baselineskip}
\titlespacing\subsection{0pt}{0.15\baselineskip}{0.08\baselineskip}
\titlespacing\subsubsection{0pt}{0.1\baselineskip}{0.08\baselineskip}

\makeatletter
\IEEEtriggercmd{\reset@font\normalfont\refsize}
\makeatother
\IEEEtriggeratref{1}

\usepackage{fancyhdr,lipsum}
\fancypagestyle{firstpage}{
  \fancyhf{}
  \fancyhead[C]{To appear at the IEEE International Joint Conference on Neural Networks (IJCNN), Maastricht, The Netherlands, June 2026.}
  \fancyfoot[C]{\thepage}
}

\begin{document}



\title{QNAS: A Neural Architecture Search Framework for Accurate and Efficient Quantum Neural Networks
\vspace{-10pt}
}

\author{\IEEEauthorblockN{Kooshan Maleki\textsuperscript{1}, Alberto Marchisio\textsuperscript{2,3}, and Muhammad Shafique\textsuperscript{2,3}\\}
\IEEEauthorblockA{
\textsuperscript{1}Department of Computer Engineering, Amirkabir University of Technology, Tehran, Iran\\
\textsuperscript{2}eBRAIN Lab, Division of Engineering, New York University Abu Dhabi (NYUAD), Abu Dhabi, UAE\\
\textsuperscript{3}Center for Quantum and Topological Systems (CQTS), NYUAD Research Institute, NYUAD, Abu Dhabi, UAE\\
Emails: kooshan.mk@gmail.com, alberto.marchisio@nyu.edu, muhammad.shafique@nyu.edu\\
}
\vspace{-30pt}
}



\maketitle
\pagestyle{empty}
\thispagestyle{empty}
\pagenumbering{gobble}
\thispagestyle{firstpage}

\begin{abstract}
Designing quantum neural networks (QNNs) that are both accurate and deployable on NISQ hardware is challenging. Handcrafted ansätze must balance expressivity, trainability, and resource use, while limited qubits often necessitate circuit cutting. Existing quantum architecture search methods primarily optimize accuracy while only heuristically controlling quantum resources (qubit count, gate counts, depth, shot complexity) and mostly ignore the exponential overhead of circuit cutting, potentially discovering architectures that are accurate but prohibitively expensive to deploy on limited qubit hardware. We introduce \emph{QNAS}, a neural architecture search framework that unifies hardware aware one shot evaluation, multi objective optimization, and cutting overhead awareness for hybrid quantum classical neural networks (HQNNs). QNAS trains a shared parameter \emph{SuperCircuit} and uses NSGA-II to optimize three objectives jointly: (i) validation error, (ii) a runtime cost proxy measuring wall clock evaluation time, and (iii) the estimated number of subcircuits under a target qubit budget. Minimizing this \emph{cutting overhead} reduces the exponential execution cost incurred by circuit partitioning. QNAS evaluates candidate HQNNs under a few epochs of training and discovers clear Pareto fronts that reveal tradeoffs between accuracy, efficiency, and cutting overhead. Across MNIST, Fashion-MNIST, and Iris benchmarks, we observe that embedding type and CNOT mode selection significantly impact both accuracy and efficiency, with angle-y embedding and sparse entangling patterns outperforming other configurations on image datasets, and amplitude embedding excelling on tabular data (Iris). On MNIST, the best architecture achieves 97.16\% test accuracy with a compact 8 qubit, 2 layer circuit; on the more challenging Fashion-MNIST, 87.38\% with a 5 qubit, 2 layer circuit; and on Iris, 100\% validation accuracy with a 4 qubit, 2 layer circuit. QNAS surfaces these design insights automatically during search, guiding practitioners toward architectures that balance accuracy, resource efficiency, and practical deployability on current hardware.
\end{abstract}

\begin{IEEEkeywords}
quantum neural networks, neural architecture search, quantum architecture search, hybrid quantum classical networks, evolutionary algorithms, NSGA-II, multi objective optimization, hardware aware design, circuit cutting, NISQ devices
\vspace{-12pt}
\end{IEEEkeywords}

\section{Introduction}
\label{Sec_Intro}

\IEEEPARstart{Q}{uantum} neural networks (QNNs) built from parameterized or variational quantum circuits (PQCs) are a promising route to near term quantum machine learning and optimization~\cite{biamonte2017quantum,cerezo2021variational,abbas2021power,killoran2019continuous,verdon2019quantum, zaman2023survey, gong2024quantum,kwak2021quantum, zaman2024comparative, innan2025next, kashif2025computational}. In practice, however, designing effective QNN architectures remains challenging. Handcrafted ansätze (e.g., hardware efficient or problem inspired circuits) must balance expressivity, trainability, and hardware efficiency and can still suffer from optimization pathologies such as barren plateaus~\cite{mcclean2018barren}, where gradients vanish exponentially with system size~\cite{larocca2025barren}; noise on Noisy Intermediate-Scale Quantum (NISQ) devices can exacerbate these effects~\cite{martyniuk2024quantum, wang2021noise, ahmed2025comparative}. Initialization strategies can partially alleviate early training plateaus~\cite{grant2019initialization}, but they do not remove the need to \emph{search} for architectures that are accurate, trainable, and compatible with hardware constraints.

Neural architecture search (NAS)~\cite{white2023neural,zoph2017neuralarchitecturesearchreinforcement} automates this design process for classical deep networks by exploring structured search spaces under tight computational budgets. NAS comprises a variety of search strategies, including reinforcement learning, evolutionary algorithms, differentiable optimization, and one shot weight sharing approaches. Orthogonally to the search methodology, NAS techniques may be single objective, typically maximizing predictive accuracy, or multi objective, where hardware-related metrics such as latency, energy consumption, or memory footprint are jointly optimized, often yielding Pareto optimal tradeoffs~\cite{sinha2024multi}. Recent work extends these ideas to quantum settings through \emph{quantum architecture search} (QAS), using evolutionary strategies, differentiable relaxations, reinforcement learning, and supercircuit weight sharing. Noise models and device connectivity are increasingly incorporated to score circuits under realistic NISQ conditions~\cite{martyniuk2024quantum,ma2024continuous, ahmed2025noisy, el2025designing}.

Despite these advances, two practical gaps limit deployability on current hardware. \emph{First}, most QAS pipelines optimize accuracy while only heuristically controlling quantum resources (qubit count, entangling gate counts, depth, and shot complexity), even though such resources strongly govern wall clock time and fidelity on real devices~\cite{ma2025understandingestimatingexecutiontime,tremba2025circuitdepthaccuratecomparing}. \emph{Second}, scaling beyond the native qubit budget requires \emph{circuit cutting}: partitioning a circuit into smaller subcircuits that can be executed on limited qubit hardware~\cite{marchisio2024cutting}. While any quantum circuit can in principle be cut, doing so incurs significant computational overhead, since each cut introduces additional measurements and classical postprocessing, and the number of circuit executions grows exponentially with the number of cuts. Existing QAS methods rarely account for this overhead, potentially discovering architectures that are accurate but prohibitively expensive to execute when cut for deployment.

\begin{figure}[t]
  \centering
  \begin{minipage}{\columnwidth}
    \centering
    \includegraphics[width=\textwidth]{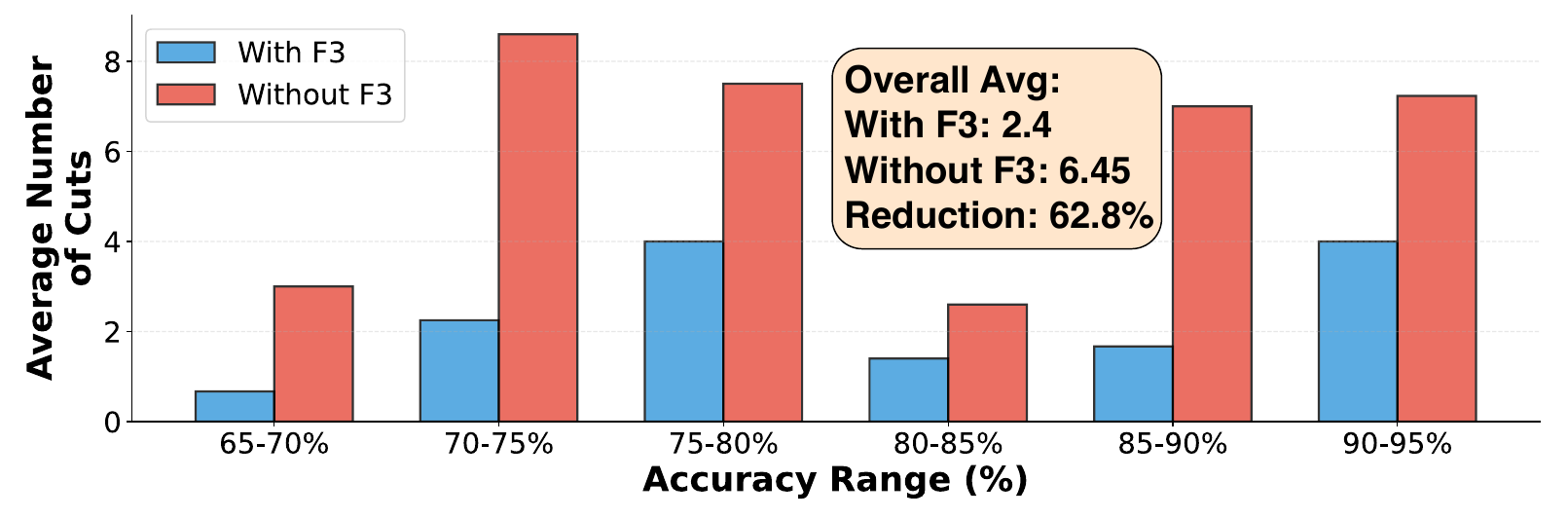}
    \vspace*{-18pt}
    \subcaption{Comparison of cutting overhead across accuracy ranges.}
    \label{fig:cuts_by_accuracy}
  \end{minipage}
  \vspace{0.5em}
  \begin{minipage}{\columnwidth}
    \centering
    \includegraphics[width=\textwidth]{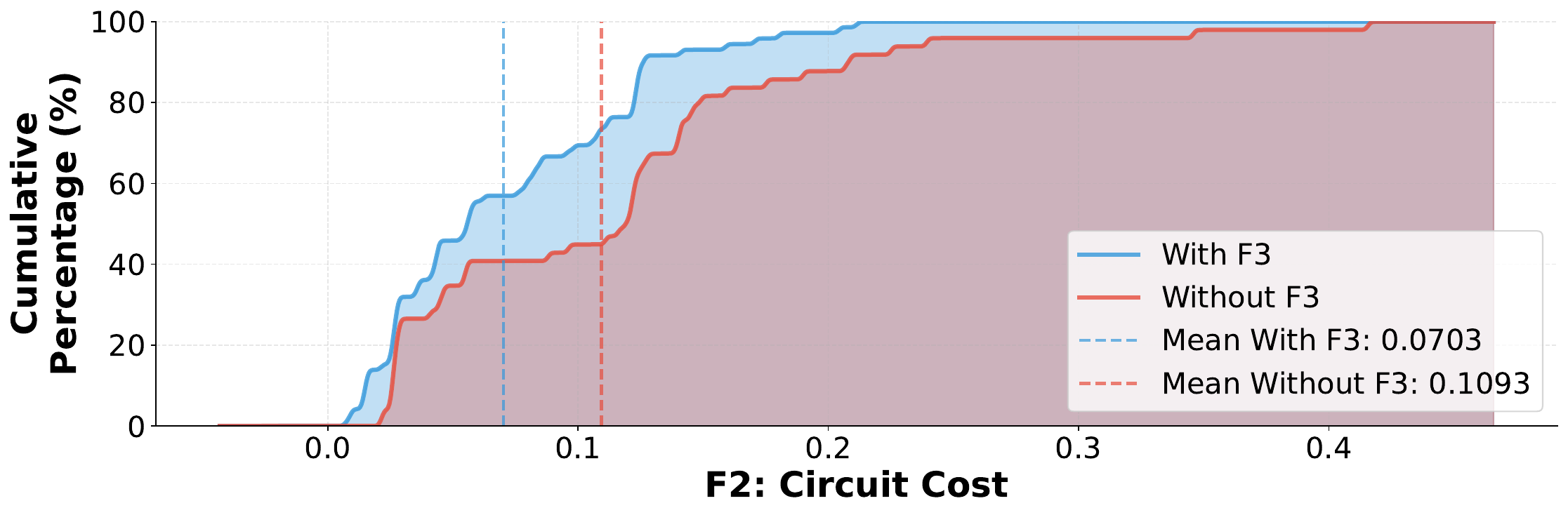}
    \vspace*{-18pt}
    \subcaption{Distribution of circuit cost (F2) for both approaches.}
    \label{fig:f2_distribution}
  \end{minipage}
  \vspace{-5pt}
  \caption{Comparison of QAS without F3 versus with F3 on MNIST: (a) Bar chart showing average number of cuts (for $Q_{\text{target}}{=}4$) across 5\% accuracy bins (solutions with accuracy $\geq$ 65\%). (b) Cumulative distribution function (CDF) of circuit cost (F2) showing percentage of solutions with F2 $\leq$ x. The ``With F3'' approach reduces both cutting overhead and circuit cost while maintaining similar accuracy.}
  \label{fig:motivational_analysis}
\end{figure}

To quantify this gap, we compare the results obtained with a multi objective QAS that does not include the cutting overhead in terms of subcircuit count (Without F3) versus a QAS where the subcircuit count is among the optimization objectives (With F3).
Fig.~\ref{fig:motivational_analysis} shows that the ``With F3'' approach reduces cutting overhead (e.g., 62.8\% on average across accuracy bins) and circuit cost (F2) while maintaining similar accuracy, indicating these objectives are complementary.

\begin{figure}[t!]
  \centering
  \includegraphics[width=\linewidth]{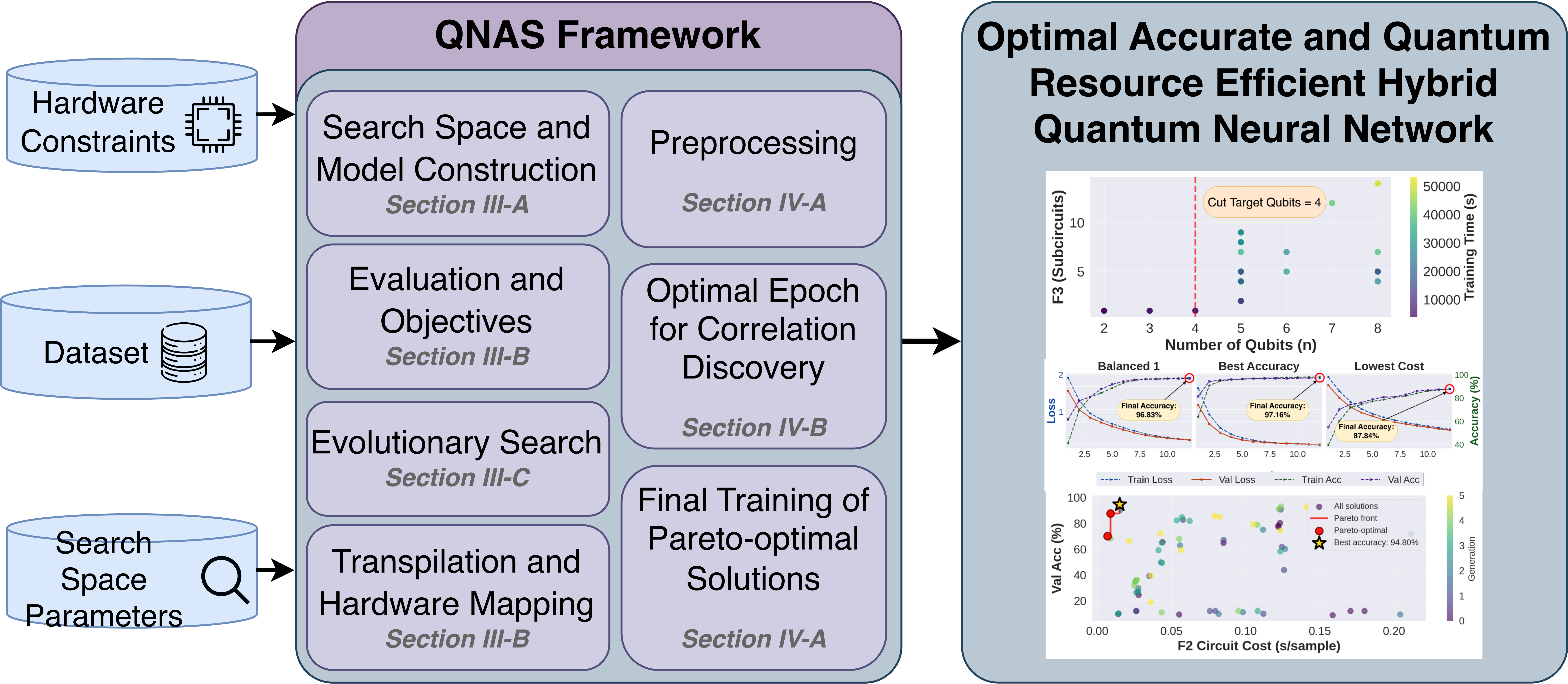}
  \caption{Overview of QNAS Framework.}
  \label{fig:qnas_overview}
\end{figure}

\textbf{This paper introduces QNAS}, a neural architecture search framework for accurate and efficient QNNs that jointly maximizes accuracy while minimizing runtime cost and cutting overhead. QNAS trains a one shot \emph{SuperCircuit} for a small number of epochs on a data subset and uses evolutionary search (NSGA-II) to optimize three objectives: (i) validation error, (ii) a \emph{circuit cost proxy} measuring wall clock execution time, and (iii) the \emph{estimated subcircuit count} under a target qubit budget, which directly governs cutting overhead. The search yields a Pareto set that balances accuracy, efficiency, and cutting overhead. Selected candidates are then retrained for more epochs on the full data for the final evaluation.
Concretely, our contributions are (see Fig.~\ref{fig:qnas_overview}):
\begin{itemize}
  \item \textbf{A hardware aware, one shot QAS pipeline.} We couple a weight sharing SuperCircuit with a few training epochs to amortize the cost of evaluating many candidates, while 
  accounting for NISQ constraints such as number of shots, circuit depth, and entangling gate usage. 
  \item \textbf{Multi objective optimization over accuracy, circuit cost, and cutting overhead.} Inspired by HW-NAS, QNAS uses NSGA-II to directly optimize a circuit cost proxy alongside an explicit subcircuit count that governs cutting overhead. 
  \item \textbf{A compact QNN search space.} QNAS searches over embedding type (angle or amplitude), qubit count, depth, entanglement range and pattern, and a small set of hyperparameters, producing architectures that are both trainable and deployable.
  \item \textbf{Empirical validation.} We evaluate QNAS on three benchmarks: MNIST, Fashion-MNIST, and Iris. On MNIST, the best discovered architecture achieves \textbf{97.16\%} test accuracy with a compact 8 qubit, 2 layer circuit; on Fashion-MNIST, \textbf{87.38\%} with a 5 qubit, 2 layer circuit; and on Iris, \textbf{100\%} validation accuracy with a 4 qubit, 2 layer circuit. Angle-y embedding with sparse entangling patterns consistently outperforms other configurations on image datasets, while amplitude embedding dominates on Iris. This demonstrates QNAS's ability to surface actionable design insights alongside Pareto optimal solutions.
\end{itemize}

\noindent \textbf{Open-Source Contribution:} To facilitate the reproduction of the experiments, and to ease the adoption of our work in the community, we release the code in a public repository at \url{https://github.com/Kooshano/QNAS}.

\noindent \textbf{Scope and implications.} QNAS does not claim to outperform the accuracy of strong classical baselines; instead, it provides a practical methodology for discovering \emph{deployable} QNN architectures under realistic constraints. By explicitly optimizing for cutting overhead alongside accuracy and runtime, QNAS helps identify architectures that execute efficiently on limited qubit devices~\cite{marchisio2024cutting}.

\noindent \textbf{Organization.} Section~\ref{Sec_Back} reviews QNNs, NAS/QAS, and circuit cutting. Section~\ref{Sec_QNAS} details the QNAS framework, objectives, and search space. Section~\ref{Sec_EvalMethod} describes datasets and the search protocol. Section~\ref{Sec_Results} reports results and analyzes tradeoffs. Section~\ref{Sec_Conclusion} concludes the paper.

\section{Background and Related Work}
\label{Sec_Back}

\begin{figure*}[!t]
  \centering
  \includegraphics[width=\textwidth]{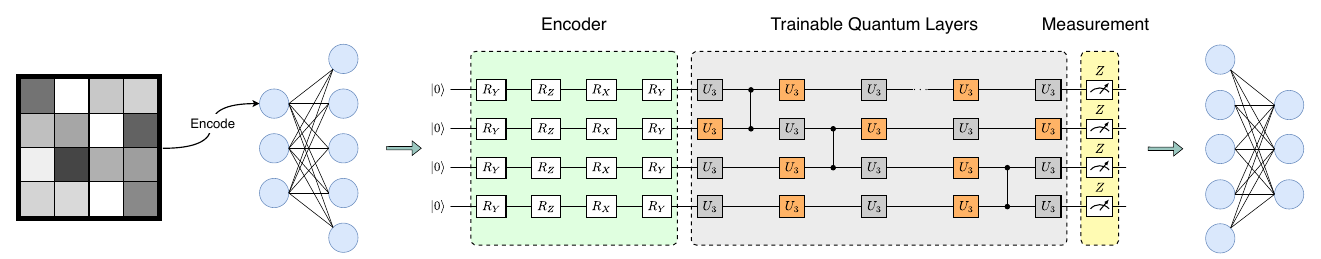} 
  \caption{Hybrid quantum classical neural network used in QNAS.}
  \label{fig:hqnn_arch}
\end{figure*}

\subsection{Variational Quantum Circuits and QNNs}
\label{Sec_HQNN_Arch}
Parameterized or variational quantum circuits (PQCs) form the backbone of variational quantum algorithms~\cite{cerezo2021variational}, including quantum neural networks (QNNs). A PQC consists of a sequence of trainable unitary gates, the \emph{ansatz}, whose parameters are optimized by a classical algorithm~\cite{zaman2024studying}. In practice, QNNs are typically realized as \emph{hybrid quantum classical neural networks} (HQNNs) that interleave classical and quantum layers to exploit quantum computation within a trainable pipeline (Fig.~\ref{fig:hqnn_arch}). The canonical HQNN has three stages:
\begin{itemize}
    \item \textbf{Classical input encoding}: a linear mapping prepares features for quantum embedding (angle or amplitude), including normalization/modulo where applicable.
    \item \textbf{Quantum variational circuit}: $L$ strongly entangling layers, each applying parameterized single qubit rotations ($R_X$, $R_Y$, $R_Z$) to every qubit followed by a set of CNOT gates; the circuit returns expectation values $\langle Z\rangle$ on all wires.
    \item \textbf{Classical output head}: a linear layer maps the $\langle Z\rangle$ vector to logits for the task loss.
\end{itemize}
This structure allows gradients to flow end to end via backpropagation using the parameter shift rule to compute quantum gradients. The design choices within each stage—embedding type, qubit count, circuit depth, and entangling pattern—define the architecture search space explored by QNAS.

The choice of ansatz influences expressivity, trainability, and hardware efficiency. Manually designed ansätze may suffer from barren plateaus where gradients vanish exponentially~\cite{mcclean2018barren,larocca2025barren}, an effect exacerbated by NISQ noise~\cite{wang2021noise}.

\subsection{Neural Architecture Search: From Classical to Quantum}

\begin{figure}
    \centering
    \includegraphics[width=1\linewidth]{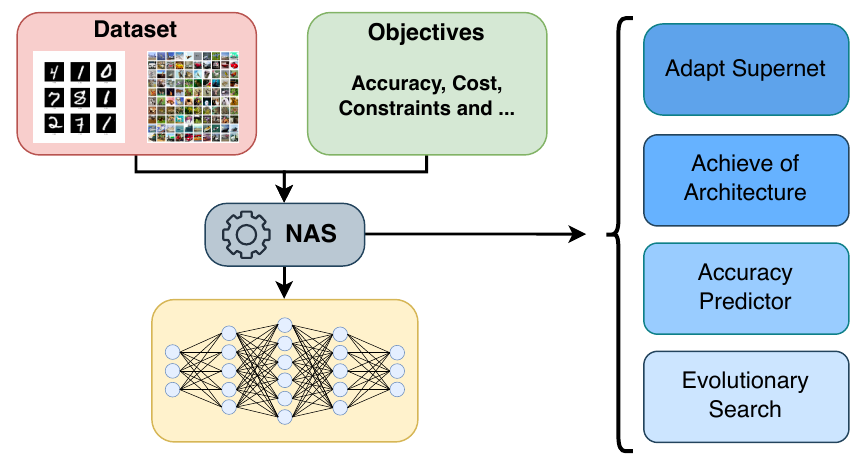}
    \caption{NAS pipeline. Data and objectives feed a NAS engine that evolves DNN architectures via a supernet and accuracy predictor.}
    \label{fig:QNAS}
\end{figure}
In classical deep learning, neural architecture search (NAS)~\cite{ren2021comprehensive,liu2021survey,elsken2019neural} automates network design by exploring architectures in a predefined search space (Fig.~\ref{fig:QNAS}). One shot NAS~\cite{guo2020single} uses a \emph{supernet} trained once; candidate networks are subnets of this supernet. Search strategies span reinforcement learning, gradient based methods, and evolutionary algorithms such as NSGA-II. Hardware aware NAS (HW-NAS)~\cite{li2025hwnasbenchhardwareawareneuralarchitecturesearch} formulates architecture design as a multi objective optimization problem that trades off accuracy against hardware cost, searching for Pareto optimal solutions~\cite{sinha2024multi, marchisio2020nascaps}.

\subsection{Quantum Architecture Search}
Quantum architecture search (QAS) extends NAS to PQCs, automatically discovering high performing quantum circuits~\cite{martyniuk2024quantum}. Modern QAS methods employ evolutionary algorithms, differentiable NAS, reinforcement learning, and Bayesian optimization~\cite{martyniuk2024quantum,kashif2025faqnas, kashif2026closing}. SuperCircuit techniques train an overparameterized circuit with shared parameters for efficient candidate evaluation~\cite{ahmed2026gat}, though parameter sharing can degrade some subcircuits' rankings. Ma \emph{et al.} address this with a continuous evolutionary framework coupling SuperCircuit training with NSGA-II~\cite{ma2024continuous}.

\subsection{Scaling QNNs with Circuit Cutting}
To scale beyond the available qubit budget, large circuits can be partitioned via \emph{circuit cutting}~\cite{marchisio2024cutting}. The overhead is significant: the number of circuit executions scales as $\mathcal{O}(4^k)$ where $k$ is the number of cuts, since each cut requires additional measurements and classical postprocessing. Given this exponential overhead, incorporating cut count as an objective in QAS is essential to steer the search toward architectures that are efficient to deploy when partitioned.

\subsection{Takeaways for QNAS}
\label{subsec:Takeaways_QNAS}
The literature suggests that a practical QNN architecture search framework should combine one shot, hardware aware evaluation (amortizing cost and scoring under realistic NISQ conditions~\cite{martyniuk2024quantum,ma2024continuous}), multi objective optimization over accuracy and quantum resource metrics~\cite{sinha2024multi,ma2024continuous}, and cutting overhead minimization (fewer subcircuits under a qubit budget reduce exponential execution cost~\cite{marchisio2024cutting}). These requirements motivate QNAS, a search framework combining one shot evaluation, multi objective evolutionary optimization, and cutting overhead awareness to discover accurate and deployable QNN architectures.
\section{The QNAS Framework}
\label{Sec_QNAS}

Our proposed Quantum Neural Architecture Search (QNAS) framework unifies hardware aware evaluation, multi objective optimization, and cutting overhead minimization to discover accurate and deployable HQNNs. Building on the HQNN structure described in Section~\ref{Sec_HQNN_Arch} and the design requirements in Section~\ref{subsec:Takeaways_QNAS}, QNAS couples few epoch training/evaluation of each candidate with NSGA-II and an explicit estimate of the subcircuit count required for deployment on limited qubit hardware.

\begin{figure}
  \centering
  \includegraphics[width=\linewidth]{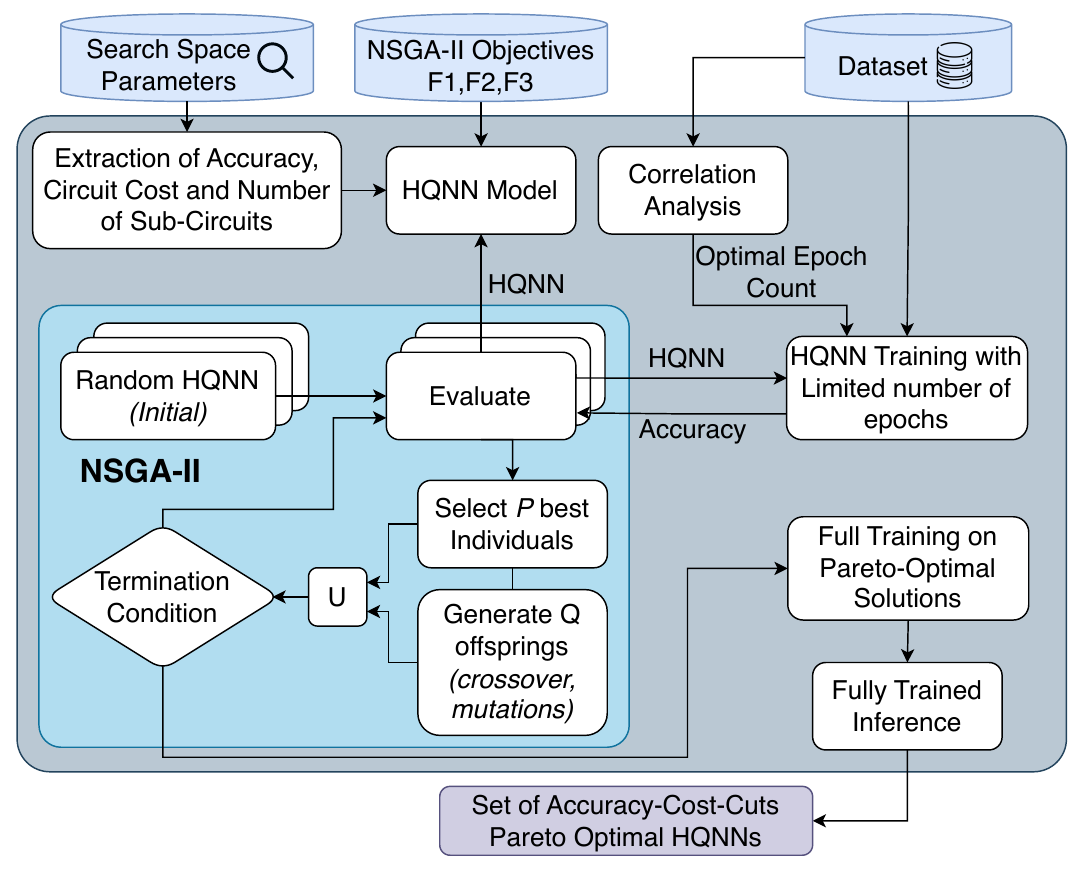}
  \caption{QNAS pipeline with NSGA-II. See Sec.~\ref{Sec_QNAS} for details.}
  \label{fig:qnas_pipeline}
\end{figure}

\subsection{Search Space and Model Construction}
\label{Sec_SearchSpace}
\begin{figure}
  \centering
  \includegraphics[width=\linewidth]{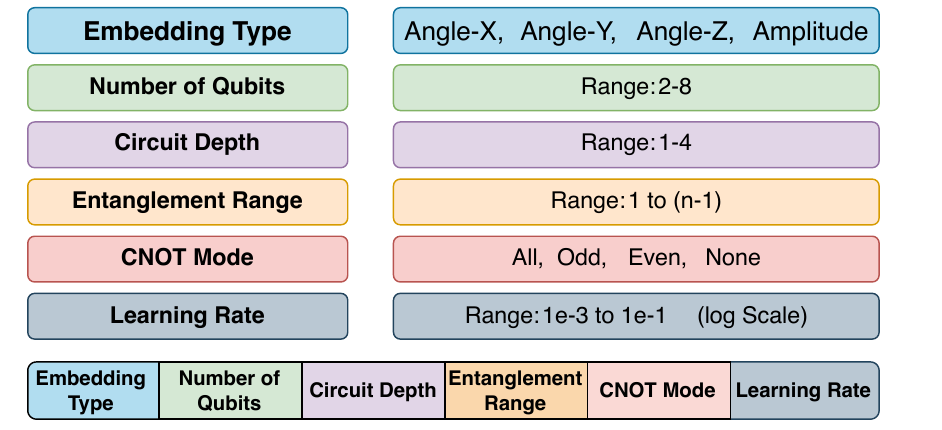}
  \caption{QNAS search space parameters and ranges. Each candidate architecture is encoded as a genome vector specifying six parameter groups: embedding type, qubit count, circuit depth, entanglement range, CNOT mode, and learning rate.}
  \label{fig:search_space}
\end{figure}

Fig.~\ref{fig:search_space} summarizes the QNAS genome and parameter ranges. Candidates instantiate the HQNN template in Fig.~\ref{fig:hqnn_arch}; the genome controls the embedding, circuit size $(n, L)$, entanglement pattern $(r_\ell, m_\ell)$, and learning rate $\eta$.

\begin{figure}[t!]
    \centering
    \begin{subfigure}[t]{0.1\textwidth}
        \centering
        \begin{adjustbox}{scale=0.45, center}
        \begin{quantikz}[row sep=0.3cm, column sep=0.25cm]
            \lstick{\scalebox{1.8}{$q_3$}} & \gate[style={fill=layerblue!80,draw=layerblue!100,thick,rounded corners=3pt}]{\scalebox{1.3}{R}} & \qw & \qw & \targ{} & \gate[style={fill=layerblue!80,draw=layerblue!100,thick,rounded corners=3pt}]{\scalebox{1.3}{R}} & \meter[style={fill=measurecolor!80,draw=black,thick,rounded corners=3pt}]{} \\
            \lstick{\scalebox{1.8}{$q_2$}} & \gate[style={fill=layerblue!80,draw=layerblue!100,thick,rounded corners=3pt}]{\scalebox{1.3}{R}} & \qw & \targ{} & \ctrl{-1} & \gate[style={fill=layerblue!80,draw=layerblue!100,thick,rounded corners=3pt}]{\scalebox{1.3}{R}} & \meter[style={fill=measurecolor!80,draw=black,thick,rounded corners=3pt}]{} \\
            \lstick{\scalebox{1.8}{$q_1$}} & \gate[style={fill=layerblue!80,draw=layerblue!100,thick,rounded corners=3pt}]{\scalebox{1.3}{R}} & \targ{} & \ctrl{-1}  & \qw & \gate[style={fill=layerblue!80,draw=layerblue!100,thick,rounded corners=3pt}]{\scalebox{1.3}{R}} & \meter[style={fill=measurecolor!80,draw=black,thick,rounded corners=3pt}]{} \\
            \lstick{\scalebox{1.8}{$q_0$}} & \gate[style={fill=layerblue!80,draw=layerblue!100,thick,rounded corners=3pt}]{\scalebox{1.3}{R}} & \ctrl{-1} & \qw & \qw & \gate[style={fill=layerblue!80,draw=layerblue!100,thick,rounded corners=3pt}]{\scalebox{1.3}{R}} & \meter[style={fill=measurecolor!80,draw=black,thick,rounded corners=3pt}]{}
        \end{quantikz}
        \end{adjustbox}
        \caption{All}
    \end{subfigure}
    \hfill
    \begin{subfigure}[t]{0.1\textwidth}
        \centering
        \begin{adjustbox}{scale=0.45, center}
        \begin{quantikz}[row sep=0.3cm, column sep=0.25cm]
            \lstick{\scalebox{1.8}{$q_3$}} & \gate[style={fill=layerblue!80,draw=layerblue!100,thick,rounded corners=3pt}]{\scalebox{1.3}{R}} & \qw & \qw & \targ{} & \gate[style={fill=layerblue!80,draw=layerblue!100,thick,rounded corners=3pt}]{\scalebox{1.3}{R}} & \meter[style={fill=measurecolor!80,draw=black,thick,rounded corners=3pt}]{} \\
            \lstick{\scalebox{1.8}{$q_2$}} & \gate[style={fill=layerblue!80,draw=layerblue!100,thick,rounded corners=3pt}]{\scalebox{1.3}{R}} & \qw & \qw & \ctrl{-1} & \gate[style={fill=layerblue!80,draw=layerblue!100,thick,rounded corners=3pt}]{\scalebox{1.3}{R}} & \meter[style={fill=measurecolor!80,draw=black,thick,rounded corners=3pt}]{} \\
            \lstick{\scalebox{1.8}{$q_1$}} & \gate[style={fill=layerblue!80,draw=layerblue!100,thick,rounded corners=3pt}]{\scalebox{1.3}{R}} & \targ{} & \qw & \qw & \gate[style={fill=layerblue!80,draw=layerblue!100,thick,rounded corners=3pt}]{\scalebox{1.3}{R}} & \meter[style={fill=measurecolor!80,draw=black,thick,rounded corners=3pt}]{} \\
            \lstick{\scalebox{1.8}{$q_0$}} & \gate[style={fill=layerblue!80,draw=layerblue!100,thick,rounded corners=3pt}]{\scalebox{1.3}{R}} & \ctrl{-1} & \qw & \qw & \gate[style={fill=layerblue!80,draw=layerblue!100,thick,rounded corners=3pt}]{\scalebox{1.3}{R}} & \meter[style={fill=measurecolor!80,draw=black,thick,rounded corners=3pt}]{}
        \end{quantikz}
        \end{adjustbox}
        \caption{Even}
    \end{subfigure}
    \hfill
    \begin{subfigure}[t]{0.1\textwidth}
        \centering
        \begin{adjustbox}{scale=0.45, center}
        \begin{quantikz}[row sep=0.3cm, column sep=0.25cm]
            \lstick{\scalebox{1.8}{$q_3$}} & \gate[style={fill=layerblue!80,draw=layerblue!100,thick,rounded corners=3pt}]{\scalebox{1.3}{R}} & \qw & \qw & \ctrl{3} & \gate[style={fill=layerblue!80,draw=layerblue!100,thick,rounded corners=3pt}]{\scalebox{1.3}{R}} & \meter[style={fill=measurecolor!80,draw=black,thick,rounded corners=3pt}]{} \\
            \lstick{\scalebox{1.8}{$q_2$}} & \gate[style={fill=layerblue!80,draw=layerblue!100,thick,rounded corners=3pt}]{\scalebox{1.3}{R}} & \targ{} & \qw & \qw & \gate[style={fill=layerblue!80,draw=layerblue!100,thick,rounded corners=3pt}]{\scalebox{1.3}{R}} & \meter[style={fill=measurecolor!80,draw=black,thick,rounded corners=3pt}]{} \\
            \lstick{\scalebox{1.8}{$q_1$}} & \gate[style={fill=layerblue!80,draw=layerblue!100,thick,rounded corners=3pt}]{\scalebox{1.3}{R}} & \ctrl{-1} & \qw & \qw & \gate[style={fill=layerblue!80,draw=layerblue!100,thick,rounded corners=3pt}]{\scalebox{1.3}{R}} & \meter[style={fill=measurecolor!80,draw=black,thick,rounded corners=3pt}]{} \\
            \lstick{\scalebox{1.8}{$q_0$}} & \gate[style={fill=layerblue!80,draw=layerblue!100,thick,rounded corners=3pt}]{\scalebox{1.3}{R}} & \qw & \qw & \targ{} & \gate[style={fill=layerblue!80,draw=layerblue!100,thick,rounded corners=3pt}]{\scalebox{1.3}{R}} & \meter[style={fill=measurecolor!80,draw=black,thick,rounded corners=3pt}]{}
        \end{quantikz}
        \end{adjustbox}
        \caption{Odd}
    \end{subfigure}
    \hfill
    \begin{subfigure}[t]{0.1\textwidth}
        \centering
        \begin{adjustbox}{scale=0.45, center}
        \begin{quantikz}[row sep=0.3cm, column sep=0.25cm]
            \lstick{\scalebox{1.8}{$q_3$}} & \gate[style={fill=layerblue!80,draw=layerblue!100,thick,rounded corners=3pt}]{\scalebox{1.3}{R}} & \qw & \qw & \qw & \qw & \gate[style={fill=layerblue!80,draw=layerblue!100,thick,rounded corners=3pt}]{\scalebox{1.3}{R}} & \meter[style={fill=measurecolor!80,draw=black,thick,rounded corners=3pt}]{} \\
            \lstick{\scalebox{1.8}{$q_2$}} & \gate[style={fill=layerblue!80,draw=layerblue!100,thick,rounded corners=3pt}]{\scalebox{1.3}{R}} & \qw & \qw & \qw & \qw & \gate[style={fill=layerblue!80,draw=layerblue!100,thick,rounded corners=3pt}]{\scalebox{1.3}{R}} & \meter[style={fill=measurecolor!80,draw=black,thick,rounded corners=3pt}]{} \\
            \lstick{\scalebox{1.8}{$q_1$}} & \gate[style={fill=layerblue!80,draw=layerblue!100,thick,rounded corners=3pt}]{\scalebox{1.3}{R}} & \qw & \qw & \qw & \qw & \gate[style={fill=layerblue!80,draw=layerblue!100,thick,rounded corners=3pt}]{\scalebox{1.3}{R}} & \meter[style={fill=measurecolor!80,draw=black,thick,rounded corners=3pt}]{} \\
            \lstick{\scalebox{1.8}{$q_0$}} & \gate[style={fill=layerblue!80,draw=layerblue!100,thick,rounded corners=3pt}]{\scalebox{1.3}{R}} & \qw & \qw & \qw & \qw & \gate[style={fill=layerblue!80,draw=layerblue!100,thick,rounded corners=3pt}]{\scalebox{1.3}{R}} & \meter[style={fill=measurecolor!80,draw=black,thick,rounded corners=3pt}]{}
        \end{quantikz}
        \end{adjustbox}
        \caption{None}
    \end{subfigure}
    \caption{CNOT mode patterns for a 4 qubit circuit with entanglement range $r{=}1$. \textbf{All}: every qubit controls its neighbor. \textbf{Odd}: only odd indexed qubits (1, 3) act as controls. \textbf{Even}: only even indexed qubits (0, 2) act as controls. \textbf{None}: no entangling gates (product state). The mode choice significantly impacts trainability and accuracy (see Sec.~\ref{Sec_Results_MNIST}).}
    \label{fig:cnot_modes}
\end{figure}
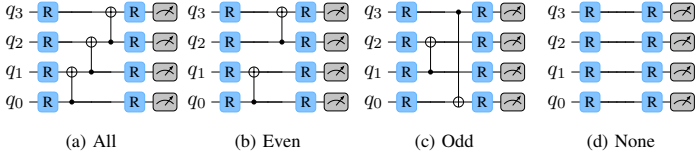

Two key entanglement parameters govern circuit structure: the \emph{entanglement range} $r_\ell$ determines how far apart qubits can be connected (Fig.~\ref{fig:ent_range}), while the \emph{CNOT mode} $m_\ell$ controls which qubits participate (Fig.~\ref{fig:cnot_modes}).

Concretely, for a given range $r$ and $n$ qubits, entangling pairs are generated as $\{(i, (i + r) \bmod n) : i \in [0, n{-}1]\}$, where qubit $i$ acts as control and qubit $(i + r) \bmod n$ as target. The modular arithmetic ensures that every qubit can potentially be entangled with every other qubit by varying $r$ across layers. E.g., $r{=}1$ yields nearest neighbor pairs, while $r{=}\lfloor n/2 \rfloor$ connects opposite qubits. The CNOT mode then filters which of these pairs are actually applied: \texttt{all} keeps every pair, \texttt{odd}/\texttt{even} retains only pairs where the control index is odd/even, and \texttt{none} removes all pairs (yielding a product state layer).

\begin{figure}
  \centering
    \begin{subfigure}[t]{0.1\textwidth}
        \centering
        \begin{adjustbox}{scale=0.4, center}
        \begin{quantikz}[row sep=0.25cm, column sep=0.2cm]
            \lstick{\scalebox{1.8}{$q_5$}} & \gate[style={fill=layerblue!80,draw=layerblue!100,thick,rounded corners=3pt}]{\scalebox{1.3}{R}} & \targ{} & \qw & \gate[style={fill=layerblue!80,draw=layerblue!100,thick,rounded corners=3pt}]{\scalebox{1.3}{R}} & \meter[style={fill=measurecolor!80,draw=black,thick,rounded corners=3pt}]{} \\
            \lstick{\scalebox{1.8}{$q_4$}} & \gate[style={fill=layerblue!80,draw=layerblue!100,thick,rounded corners=3pt}]{\scalebox{1.3}{R}} & \ctrl{-1} & \targ{} & \gate[style={fill=layerblue!80,draw=layerblue!100,thick,rounded corners=3pt}]{\scalebox{1.3}{R}} & \meter[style={fill=measurecolor!80,draw=black,thick,rounded corners=3pt}]{} \\
            \lstick{\scalebox{1.8}{$q_3$}} & \gate[style={fill=layerblue!80,draw=layerblue!100,thick,rounded corners=3pt}]{\scalebox{1.3}{R}} & \targ{} & \ctrl{-1} & \gate[style={fill=layerblue!80,draw=layerblue!100,thick,rounded corners=3pt}]{\scalebox{1.3}{R}} & \meter[style={fill=measurecolor!80,draw=black,thick,rounded corners=3pt}]{} \\
            \lstick{\scalebox{1.8}{$q_2$}} & \gate[style={fill=layerblue!80,draw=layerblue!100,thick,rounded corners=3pt}]{\scalebox{1.3}{R}} & \ctrl{-1} & \targ{} & \gate[style={fill=layerblue!80,draw=layerblue!100,thick,rounded corners=3pt}]{\scalebox{1.3}{R}} & \meter[style={fill=measurecolor!80,draw=black,thick,rounded corners=3pt}]{} \\
            \lstick{\scalebox{1.8}{$q_1$}} & \gate[style={fill=layerblue!80,draw=layerblue!100,thick,rounded corners=3pt}]{\scalebox{1.3}{R}} & \targ{} & \ctrl{-1} & \gate[style={fill=layerblue!80,draw=layerblue!100,thick,rounded corners=3pt}]{\scalebox{1.3}{R}} & \meter[style={fill=measurecolor!80,draw=black,thick,rounded corners=3pt}]{} \\
            \lstick{\scalebox{1.8}{$q_0$}} & \gate[style={fill=layerblue!80,draw=layerblue!100,thick,rounded corners=3pt}]{\scalebox{1.3}{R}} & \ctrl{-1} & \qw & \gate[style={fill=layerblue!80,draw=layerblue!100,thick,rounded corners=3pt}]{\scalebox{1.3}{R}} & \meter[style={fill=measurecolor!80,draw=black,thick,rounded corners=3pt}]{}
        \end{quantikz}
        \end{adjustbox}
        \caption{$r=1$}
    \end{subfigure}
    \hfill
    \begin{subfigure}[t]{0.1\textwidth}
        \centering
        \begin{adjustbox}{scale=0.4, center}
        \begin{quantikz}[row sep=0.25cm, column sep=0.2cm]
            \lstick{\scalebox{1.8}{$q_5$}} & \gate[style={fill=layerblue!80,draw=layerblue!100,thick,rounded corners=3pt}]{\scalebox{1.3}{R}} & \targ{} & \qw & \qw & \qw & \gate[style={fill=layerblue!80,draw=layerblue!100,thick,rounded corners=3pt}]{\scalebox{1.3}{R}} & \meter[style={fill=measurecolor!80,draw=black,thick,rounded corners=3pt}]{} \\
            \lstick{\scalebox{1.8}{$q_4$}} & \gate[style={fill=layerblue!80,draw=layerblue!100,thick,rounded corners=3pt}]{\scalebox{1.3}{R}} & \qw & \targ{} & \qw & \qw & \gate[style={fill=layerblue!80,draw=layerblue!100,thick,rounded corners=3pt}]{\scalebox{1.3}{R}} & \meter[style={fill=measurecolor!80,draw=black,thick,rounded corners=3pt}]{} \\
            \lstick{\scalebox{1.8}{$q_3$}} & \gate[style={fill=layerblue!80,draw=layerblue!100,thick,rounded corners=3pt}]{\scalebox{1.3}{R}} & \ctrl{-2} & \qw & \targ{} & \qw & \gate[style={fill=layerblue!80,draw=layerblue!100,thick,rounded corners=3pt}]{\scalebox{1.3}{R}} & \meter[style={fill=measurecolor!80,draw=black,thick,rounded corners=3pt}]{} \\
            \lstick{\scalebox{1.8}{$q_2$}} & \gate[style={fill=layerblue!80,draw=layerblue!100,thick,rounded corners=3pt}]{\scalebox{1.3}{R}} & \qw & \ctrl{-2} & \qw & \targ{} & \gate[style={fill=layerblue!80,draw=layerblue!100,thick,rounded corners=3pt}]{\scalebox{1.3}{R}} & \meter[style={fill=measurecolor!80,draw=black,thick,rounded corners=3pt}]{} \\
            \lstick{\scalebox{1.8}{$q_1$}} & \gate[style={fill=layerblue!80,draw=layerblue!100,thick,rounded corners=3pt}]{\scalebox{1.3}{R}} & \qw & \qw & \ctrl{-2} & \qw & \gate[style={fill=layerblue!80,draw=layerblue!100,thick,rounded corners=3pt}]{\scalebox{1.3}{R}} & \meter[style={fill=measurecolor!80,draw=black,thick,rounded corners=3pt}]{} \\
            \lstick{\scalebox{1.8}{$q_0$}} & \gate[style={fill=layerblue!80,draw=layerblue!100,thick,rounded corners=3pt}]{\scalebox{1.3}{R}} & \qw & \qw & \qw & \ctrl{-2} & \gate[style={fill=layerblue!80,draw=layerblue!100,thick,rounded corners=3pt}]{\scalebox{1.3}{R}} & \meter[style={fill=measurecolor!80,draw=black,thick,rounded corners=3pt}]{}
        \end{quantikz}
        \end{adjustbox}
        \caption{$r=2$}
    \end{subfigure}
    \hfill
    \begin{subfigure}[t]{0.1\textwidth}
        \centering
        \begin{adjustbox}{scale=0.4, center}
        \begin{quantikz}[row sep=0.25cm, column sep=0.2cm]
            \lstick{\scalebox{1.8}{$q_5$}} & \gate[style={fill=layerblue!80,draw=layerblue!100,thick,rounded corners=3pt}]{\scalebox{1.3}{R}} & \targ{} & \qw & \qw & \gate[style={fill=layerblue!80,draw=layerblue!100,thick,rounded corners=3pt}]{\scalebox{1.3}{R}} & \meter[style={fill=measurecolor!80,draw=black,thick,rounded corners=3pt}]{} \\
            \lstick{\scalebox{1.8}{$q_4$}} & \gate[style={fill=layerblue!80,draw=layerblue!100,thick,rounded corners=3pt}]{\scalebox{1.3}{R}} & \qw & \targ{} & \qw & \gate[style={fill=layerblue!80,draw=layerblue!100,thick,rounded corners=3pt}]{\scalebox{1.3}{R}} & \meter[style={fill=measurecolor!80,draw=black,thick,rounded corners=3pt}]{} \\
            \lstick{\scalebox{1.8}{$q_3$}} & \gate[style={fill=layerblue!80,draw=layerblue!100,thick,rounded corners=3pt}]{\scalebox{1.3}{R}} & \qw & \qw & \targ{} & \gate[style={fill=layerblue!80,draw=layerblue!100,thick,rounded corners=3pt}]{\scalebox{1.3}{R}} & \meter[style={fill=measurecolor!80,draw=black,thick,rounded corners=3pt}]{} \\
            \lstick{\scalebox{1.8}{$q_2$}} & \gate[style={fill=layerblue!80,draw=layerblue!100,thick,rounded corners=3pt}]{\scalebox{1.3}{R}} & \ctrl{-3} & \qw & \qw & \gate[style={fill=layerblue!80,draw=layerblue!100,thick,rounded corners=3pt}]{\scalebox{1.3}{R}} & \meter[style={fill=measurecolor!80,draw=black,thick,rounded corners=3pt}]{} \\
            \lstick{\scalebox{1.8}{$q_1$}} & \gate[style={fill=layerblue!80,draw=layerblue!100,thick,rounded corners=3pt}]{\scalebox{1.3}{R}} & \qw & \ctrl{-3} & \qw & \gate[style={fill=layerblue!80,draw=layerblue!100,thick,rounded corners=3pt}]{\scalebox{1.3}{R}} & \meter[style={fill=measurecolor!80,draw=black,thick,rounded corners=3pt}]{} \\
            \lstick{\scalebox{1.8}{$q_0$}} & \gate[style={fill=layerblue!80,draw=layerblue!100,thick,rounded corners=3pt}]{\scalebox{1.3}{R}} & \qw & \qw & \ctrl{-3} & \gate[style={fill=layerblue!80,draw=layerblue!100,thick,rounded corners=3pt}]{\scalebox{1.3}{R}} & \meter[style={fill=measurecolor!80,draw=black,thick,rounded corners=3pt}]{}
        \end{quantikz}
        \end{adjustbox}
        \caption{$r=3$}
    \end{subfigure}
    \hfill
    \begin{subfigure}[t]{0.1\textwidth}
        \centering
        \begin{adjustbox}{scale=0.4, center}
        \begin{quantikz}[row sep=0.25cm, column sep=0.2cm]
            \lstick{\scalebox{1.8}{$q_5$}} & \gate[style={fill=layerblue!80,draw=layerblue!100,thick,rounded corners=3pt}]{\scalebox{1.3}{R}} & \targ{} & \qw & \gate[style={fill=layerblue!80,draw=layerblue!100,thick,rounded corners=3pt}]{\scalebox{1.3}{R}} & \meter[style={fill=measurecolor!80,draw=black,thick,rounded corners=3pt}]{} \\
            \lstick{\scalebox{1.8}{$q_4$}} & \gate[style={fill=layerblue!80,draw=layerblue!100,thick,rounded corners=3pt}]{\scalebox{1.3}{R}} & \qw & \targ{} & \gate[style={fill=layerblue!80,draw=layerblue!100,thick,rounded corners=3pt}]{\scalebox{1.3}{R}} & \meter[style={fill=measurecolor!80,draw=black,thick,rounded corners=3pt}]{} \\
            \lstick{\scalebox{1.8}{$q_3$}} & \gate[style={fill=layerblue!80,draw=layerblue!100,thick,rounded corners=3pt}]{\scalebox{1.3}{R}} & \qw & \qw & \gate[style={fill=layerblue!80,draw=layerblue!100,thick,rounded corners=3pt}]{\scalebox{1.3}{R}} & \meter[style={fill=measurecolor!80,draw=black,thick,rounded corners=3pt}]{} \\
            \lstick{\scalebox{1.8}{$q_2$}} & \gate[style={fill=layerblue!80,draw=layerblue!100,thick,rounded corners=3pt}]{\scalebox{1.3}{R}} & \qw & \qw & \gate[style={fill=layerblue!80,draw=layerblue!100,thick,rounded corners=3pt}]{\scalebox{1.3}{R}} & \meter[style={fill=measurecolor!80,draw=black,thick,rounded corners=3pt}]{} \\
            \lstick{\scalebox{1.8}{$q_1$}} & \gate[style={fill=layerblue!80,draw=layerblue!100,thick,rounded corners=3pt}]{\scalebox{1.3}{R}} & \ctrl{-4} & \qw & \gate[style={fill=layerblue!80,draw=layerblue!100,thick,rounded corners=3pt}]{\scalebox{1.3}{R}} & \meter[style={fill=measurecolor!80,draw=black,thick,rounded corners=3pt}]{} \\
            \lstick{\scalebox{1.8}{$q_0$}} & \gate[style={fill=layerblue!80,draw=layerblue!100,thick,rounded corners=3pt}]{\scalebox{1.3}{R}} & \qw & \ctrl{-4} & \gate[style={fill=layerblue!80,draw=layerblue!100,thick,rounded corners=3pt}]{\scalebox{1.3}{R}} & \meter[style={fill=measurecolor!80,draw=black,thick,rounded corners=3pt}]{}
        \end{quantikz}
        \end{adjustbox}
        \caption{$r=4$}
    \end{subfigure}

    \caption{Effect of entanglement range $r$ on CNOT connectivity for a 6 qubit circuit. $r=1$: nearest neighbor ($q_i \rightarrow q_{i+1}$). $r=2$: skip one ($q_i \rightarrow q_{i+2}$). $r=3$: opposite pairs ($q_i \rightarrow q_{i+3}$). $r=4$: skip three ($q_i \rightarrow q_{i+4}$). Larger ranges enable longer distance correlations but may increase hardware routing overhead.}
  \label{fig:ent_range}
\end{figure}
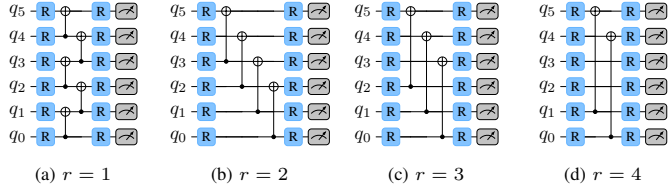

\subsection{Evaluation and Objectives}
For efficient scoring, QNAS trains each HQNN briefly on a subset
and evaluates three objectives:
\begin{itemize}
  \item \textbf{Accuracy loss:}
  \begin{equation}
    F_{1} = 1 - \mathrm{Accuracy}_{\text{val}},
  \end{equation}
  where $\mathrm{Accuracy}_{\text{val}}$ is measured on a held out validation set.
  \item \textbf{Runtime cost proxy (implementation aligned):}
\begin{equation}
  F_{2} \;=\; \frac{t_{\mathrm{val}}}{N_{\mathrm{val}}}
  \quad \text{(seconds/sample)},
\end{equation}
where \(t_{\mathrm{val}}\) is the GPU synchronized wall clock time for the validation loop and \(N_{\mathrm{val}}\) is the number of validation samples; this aligns \(F_2\) with effective execution cost under the chosen backend.
%
%
  \item \textbf{Cutting overhead (subcircuit count):}
\begin{multline}
  F_{3} = 
  \text{Estimated number of subcircuits required} \\
  \text{under a target qubit budget } Q_{\text{target}},
\end{multline}
  computed via circuit cut placement heuristics or a conservative fallback. Minimizing $F_3$ is critical because circuit cutting incurs exponential overhead: each cut requires additional measurements and classical postprocessing, and the total number of circuit executions scales as $\mathcal{O}(4^k)$ where $k$ is the number of cuts~\cite{marchisio2024cutting}. Architectures with fewer required subcircuits are thus far more efficient to deploy on limited qubit hardware.
\end{itemize}

\begin{figure}[t!]
  \centering
  \includegraphics[width=\linewidth]{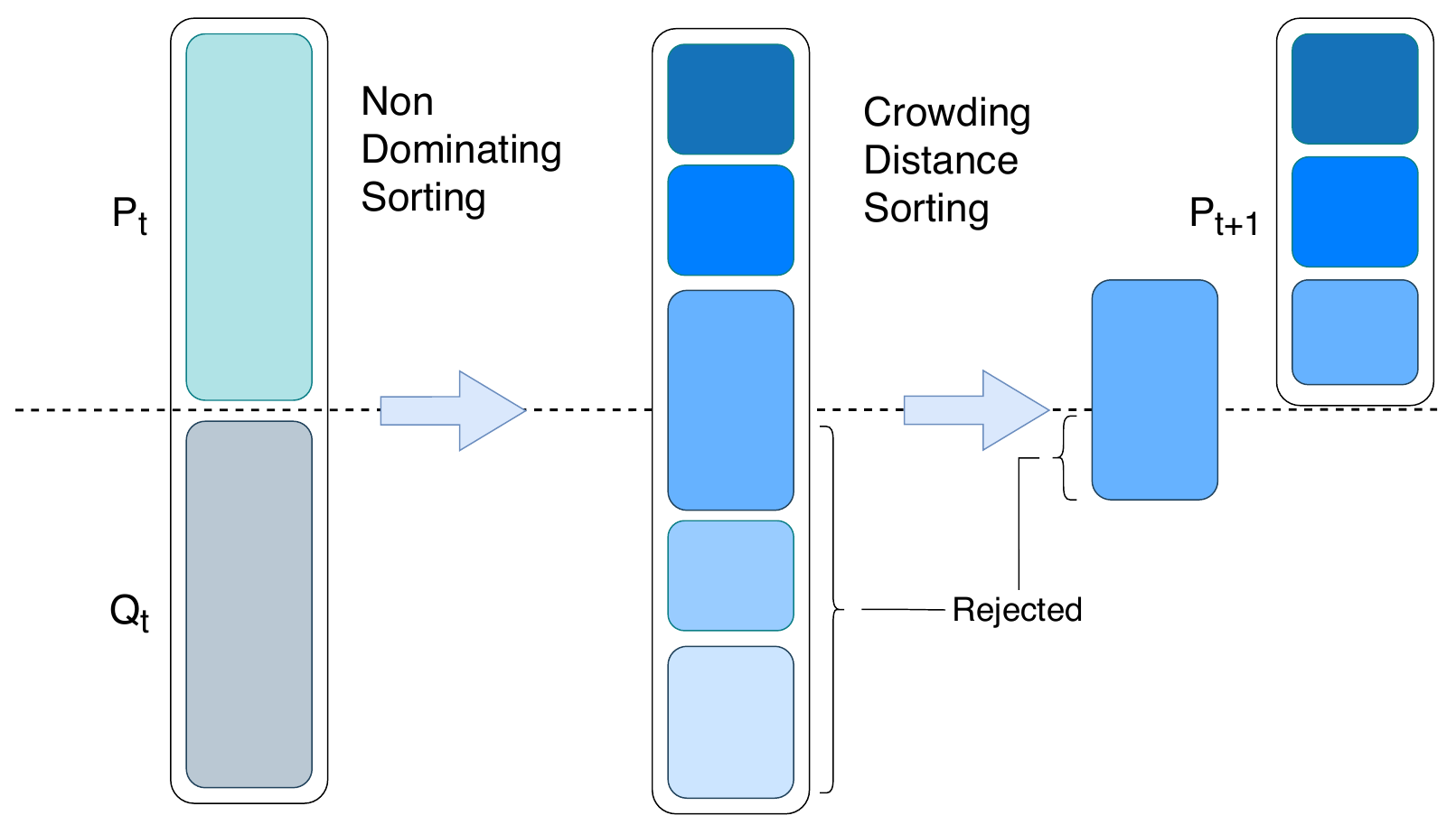}
  \caption{NSGA-II selection mechanism. \textbf{Left}: Parent population $P_t$. \textbf{Middle}: Combined population $P_t \cup Q_t$ sorted into non dominated fronts (darker = better front). \textbf{Right}: Selection of $P_{t+1}$ from top fronts; the dashed line indicates where crowding distance breaks ties within a partially selected front.}
  \label{fig:nsga2}
\end{figure}

\subsection{Evolutionary Search}
QNAS uses the Non-dominated Sorting Genetic Algorithm II (NSGA-II)~\cite{deb2002fast} to navigate the multi objective search space. Fig.~\ref{fig:nsga2} illustrates the core mechanism: at each generation, the parent population $P_t$ is combined with offspring $Q_t$ (generated via crossover and mutation). The combined population is then sorted into \emph{non dominated fronts}, groups of solutions where no solution dominates another within the same front~\cite{pallasdies2021neural,van1998evolutionary,panichella2022improved,li2022hierarchy}. The first front (darkest shade) contains Pareto optimal solutions; subsequent fronts contain progressively dominated solutions. Selection fills the next generation $P_{t+1}$ by taking complete fronts in order; when a front would exceed the population size, crowding distance is used to prefer solutions in sparse regions of the objective space, maintaining diversity.

From a random initial population, QNAS applies SBX crossover and polynomial mutation to generate offspring, evaluates $(F_1,F_2,F_3)$ for each genome, and iterates for a fixed number of generations.
The search yields a Pareto set balancing accuracy, cost, and cutting overhead. 
A selected model (typically the highest accuracy point) is then retrained for more epochs on full data for final reporting.

\subsection{Transpilation and Hardware Mapping}
\label{Sec_Transpilation}

To assess deployability on specific NISQ devices, QNAS supports a standard \emph{transpilation} step that maps each candidate variational circuit from an ideal, device-agnostic description to a device-constrained implementation~\cite{kremer2024practical}. Given a candidate circuit $\mathcal{C}$ specified by $(n,L,\{r_\ell\},\{m_\ell\})$ and a target backend characterized by a coupling graph $G=(V,E)$ and a native basis gate set $\mathcal{B}$, transpilation~\cite{hua2023qasmtrans} produces an executable circuit
\begin{equation}
  \widetilde{\mathcal{C}}
  \;=\;
  \mathrm{Transpile}\!\left(\mathcal{C}\,;\,G,\mathcal{B},\mathcal{O}\right),
\end{equation}
where $\mathcal{O}$ denotes routing and optimization options. In general, transpilation can increase effective depth and two qubit gate count due to (i) routing overhead (e.g., SWAP insertion when the entangling pattern implied by $r_\ell$ conflicts with $G$) and (ii) basis decomposition, both affecting runtime and fidelity.

\textbf{Snapshot based backends for noise aware simulation:}
QNAS can employ snapshot based backends (``fake backends'') that mimic IBM Quantum systems using historical snapshots of device connectivity, basis gates, and qubit-level noise properties (e.g., $T_1$, $T_2$, gate error rates, readout errors). Crucially, these backends include calibrated noise models derived from real hardware characterization data, enabling noisy simulation that reflects realistic NISQ device behavior. This means that circuit executions incorporate depolarizing errors, thermal relaxation, and measurement noise representative of actual quantum processors. As a result, architectures evaluated using fake backends are assessed under conditions that approximate real hardware execution, ensuring that reported accuracies account for noise-induced performance degradation typical of current NISQ devices.

\textbf{Deployment oriented reporting:}
Transpilation provides deployment centric diagnostics complementing the QNAS objectives, including transpiled depth $\widetilde{L}$, two qubit gate count $\widetilde{N}_{2q}$, and routing overhead (e.g., SWAP count). These metrics quantify how a candidate's entanglement design interacts with hardware topology.

\textbf{Relationship to the runtime objective:}
When transpilation is enabled, the runtime objective $F_2$ can be measured on the transpiled circuit $\widetilde{\mathcal{C}}$, thereby incorporating compilation and routing overhead into the tradeoffs produced by QNAS.

\section{Evaluation Methodology}
\label{Sec_EvalMethod}

\subsection{Tasks, Datasets, and Preprocessing}
QNAS is designed for supervised classification tasks and can be applied to any dataset with standard train/validation/test splits. During the \emph{search phase}, the training set is used for parameter optimization and a held out validation set (e.g., 10\% of training data) provides fitness scores $(F_1, F_2, F_3)$ for NSGA-II selection. The test set is reserved exclusively for \emph{final evaluation} of Pareto optimal architectures after retraining.

\textbf{Preprocessing:} Input features are scaled to $[0,1]$. For high-dimensional inputs (e.g., images), dimensionality reduction via PCA extracts the top $n$ principal components matching the qubit count. For \emph{angle embeddings}, these $n$ features are standardized to zero mean and unit variance, then encoded as rotation angles $\theta_i \in [0, 2\pi)$. For \emph{amplitude embeddings}, feature vectors are $\ell_2$ normalized and zero padded to length $2^n$ to match the Hilbert space dimension.

\textbf{Instantiation:} In this work, we evaluate QNAS on three benchmarks: MNIST~\cite{6296535} and Fashion-MNIST~\cite{xiao2017fashionmnistnovelimagedataset}, each with 60{,}000 training / 10{,}000 test grayscale images across 10 classes (Fashion-MNIST is more challenging due to higher intra-class variation); and the Iris dataset~\cite{iris_53} (3-class classification with 4-dimensional features). For MNIST and Fashion-MNIST, images are flattened to 784-dimensional vectors before PCA reduction to $n$ features. The target qubit budget $Q_{\text{target}}$ is set per dataset to reflect realistic hardware constraints: $Q_{\text{target}}{=}4$ for MNIST and Fashion-MNIST (moderate-scale image tasks), and $Q_{\text{target}}{=}2$ for Iris (small-scale tabular task).

\subsection{Epoch Selection via Correlation Analysis}
\label{Sec_EpochCorrelation}

A key design question for QNAS is: \emph{how many training epochs are needed during NSGA-II search to reliably rank candidate architectures?} Training each candidate for many epochs would yield accurate fitness estimates but is computationally prohibitive when evaluating hundreds of candidates. Conversely, too few epochs may produce rankings that do not correlate with final model performance, leading to suboptimal architecture selection.

To answer this question, we conduct a correlation study between checkpoint accuracies at various epochs and final (fully trained) accuracy. We train a diverse set of randomly sampled architectures for several epochs each and record validation accuracy at multiple checkpoints. We then compute both Pearson ($r$) and Spearman ($\rho$) correlation coefficients between each checkpoint's accuracy and the final accuracy.

Our analysis reveals that strong correlation ($r{>}0.8$) is typically achieved within the first few epochs, with diminishing returns thereafter. Few epoch training provides a reasonably accurate proxy for final model performance, enabling efficient candidate screening during evolutionary search while significantly reducing computational cost. Based on this analysis, QNAS uses few epoch training during search, and Pareto optimal architectures from the search phase are then retrained for additional epochs on full data for final performance evaluation.

\section{Results and Discussion}
\label{Sec_Results}

\subsection{Results on MNIST}
\label{Sec_Results_MNIST}

\begin{figure}
  \centering
  \includegraphics[width=\linewidth]{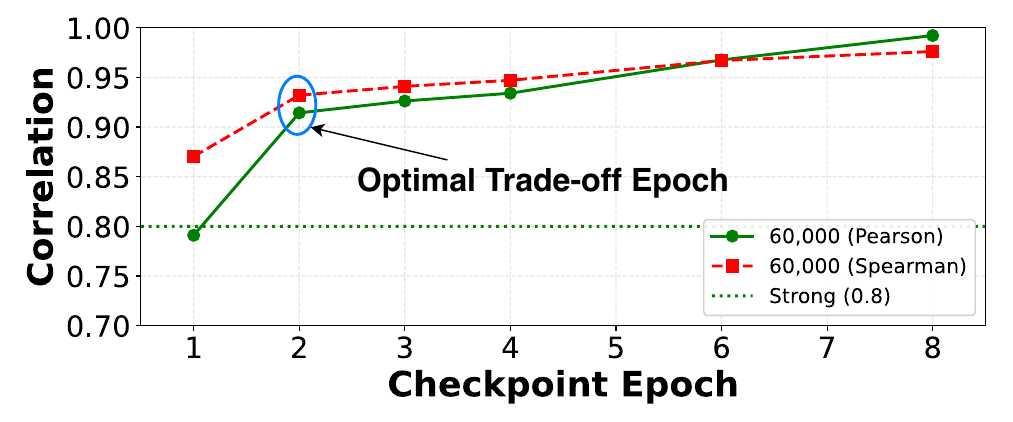}
  \caption{Correlation between checkpoint accuracy and final (epoch~10) accuracy on MNIST using the full 60{,}000 training set. The dashed line marks $r{=}0.8$ (``strong'' correlation). By epoch~2, correlations exceed 0.9, validating short training budgets during search.}
  \label{fig:epoch_correlation}
\end{figure}

\textbf{Epoch selection:}
To validate the few epoch training strategy (Sec.~\ref{Sec_EpochCorrelation}) on MNIST, we trained 24 diverse randomly sampled architectures for 10 epochs each on the full 60{,}000 samples of the training set and computed correlations between checkpoint and final accuracy. As shown in Fig.~\ref{fig:epoch_correlation}, correlations are moderate at epoch~1 but jump dramatically by epoch~2, with both Pearson and Spearman coefficients exceeding 0.9. This validates the use of 2 epoch training during search, reducing computational cost by $5{\times}$ while maintaining high fidelity architecture rankings.

\textbf{Search setup and budget:}
Table~\ref{tab:search_space} shows the MNIST search space instantiation, yielding approximately $10^4$--$10^6$ configurations depending on depth.

\begin{table}[ht]
\caption{MNIST search space configuration.}
\label{tab:search_space}
\centering
\small
\resizebox{\linewidth}{!}{%
\begin{tabular}{ll}
\toprule
\textbf{Parameter} & \textbf{Range} \\
\midrule
Embedding type & angle-$X$, angle-$Y$, angle-$Z$, amplitude \\
Qubits $n$ & $[2, 8]$ \\
Depth $L$ & $[1, 4]$ \\
Entanglement range $r$ (per layer) & $[1, n{-}1]$ \\
CNOT mode (per layer) & all, odd, even, none \\
Learning rate $\eta$ & $[10^{-3}, 5{\times}10^{-3}]$ \\
\bottomrule
\end{tabular}
}
\end{table}

Over six generations, the NSGA-II search evaluated \textbf{72} HQNN candidates using 2 epochs of training per candidate on a GPU backend (\texttt{lightning.gpu}). 
We report validation accuracy (in \%) along with $F_2$ (runtime cost in s/sample) and $F_3$ (subcircuit count with $Q_{\text{target}}{=}4$). 
A compact summary is provided in Table~\ref{tab:mnist_summary}. Over six generations, median and highest accuracy improved steadily, while runtime cost decreased.

\begin{table}
\caption{MNIST search phase summary.}
\label{tab:mnist_summary}
\centering
\small
\setlength{\tabcolsep}{3pt}
\resizebox{\linewidth}{!}{%
\begin{tabular}{llll}
\toprule
                        Metric &               Value &                        Metric &               Value \\
\midrule
         Total evaluations &                   72 & Best search val accuracy (\%) &               94.80 \\
                Generations &                    6 & Best final test accuracy (\%) &               97.16 \\
         Training epochs/eval &                    2 &   Mean $\pm$ std accuracy (\%) &       51.32 $\pm$ 29.84 \\
         $Q_{\text{target}}$ (qubits) &                    4 &      Mean $F_2$ (s/sample) &               0.070 \\
      Total search time (h) &               71.19 & & \\
\bottomrule
\end{tabular}%
}
\end{table}

\begin{figure}[t!]
  \centering
  \includegraphics[width=\linewidth]{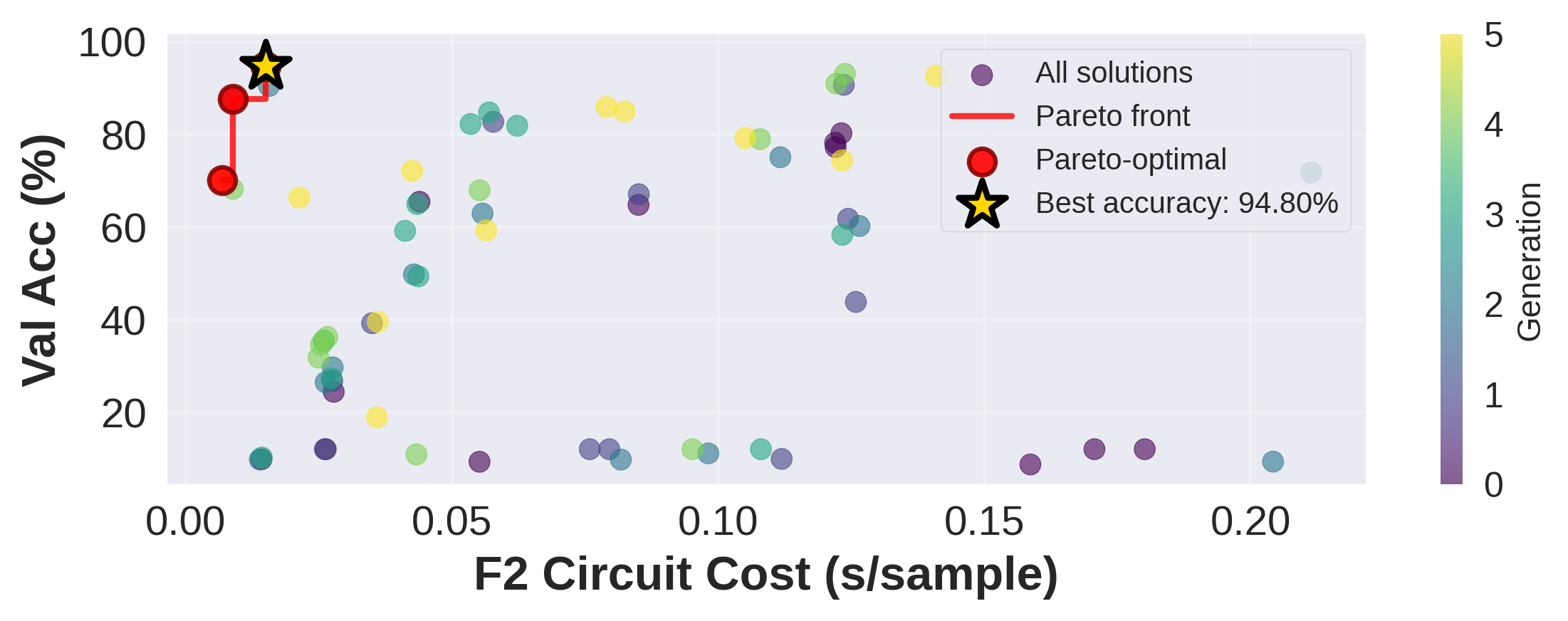}
  \caption{Pareto tradeoff between validation accuracy and runtime cost ($F_2$). The star marks the best accuracy (94.80\%).}
  \label{fig:mnist_pareto}
\end{figure}

\textbf{Pareto tradeoffs and selected architectures:}
The tradeoff between accuracy and runtime cost is visualized in Fig.~\ref{fig:mnist_pareto}; 
Pareto efficient points (minimizing $F_2$ and maximizing accuracy) form a clear front. 
The best search phase model achieved \textbf{94.80\%} validation accuracy (generation~4), 
with configuration 8q-2d angle-y, CNOT mode none-odd, $F_2{=}0.0150$\,s/sample, $F_3{=}5$. 
Table~\ref{tab:mnist_results} lists the top configurations and Pareto optimal solutions over $(F_1, F_2, F_3)$.

\begin{table}[t!]
\caption{Search results: top configurations by accuracy and Pareto optimal solutions. Rows marked $\bullet$ are Pareto optimal over $(F_1, F_2, F_3)$; unmarked rows are top accuracy configurations. Angle-y with CNOT mode ``none-odd'' dominates both categories.}
\label{tab:mnist_results}
\centering
\small
\setlength{\tabcolsep}{2.5pt}
\resizebox{\linewidth}{!}{%
\begin{tabular}{clrrrllrrr}
\toprule
 & EvalID & Gen &  Embed &  Qubits &  Depth &  CNOT Mode &  ValAcc(\%) &  $F_2$ (s/samp) &  $F_3$ \\
\midrule
\multicolumn{10}{l}{\textit{Top accuracy configurations}} \\
$\bullet$ & 45 &   4 & angle-y &       8 &      2 &  none-odd &       94.80 &  0.0150 &   5 \\
          & 51 &   5 & angle-y &       8 &      2 &  none-odd &       94.17 &  0.0151 &   5 \\
          & 70 &   6 & angle-y &       8 &      2 &  none-odd &       93.24 &  0.0152 &   5 \\
$\bullet$ & 49 &   5 & angle-y &       5 &      2 &  none-odd &       93.12 &  0.1238 &   4 \\
          & 66 &   6 & angle-y &       8 &      1 &  odd      &       92.65 &  0.1409 &   4 \\
          & 57 &   5 & angle-x &       5 &      2 &  odd-odd  &       91.02 &  0.1222 &   5 \\
          & 15 &   2 & angle-x &       5 &      2 &  odd-odd  &       90.75 &  0.1236 &   5 \\
          & 30 &   3 & angle-x &       8 &      2 &  odd-odd  &       90.50 &  0.0156 &   7 \\
$\bullet$ & 24 &   2 & angle-y &       5 &      2 &  none-odd &       87.72 &  0.0089 &   4 \\
          & 65 &   6 & angle-y &       5 &      1 &  none     &       85.94 &  0.0790 &   2 \\
\midrule
\multicolumn{10}{l}{\textit{Additional Pareto optimal solutions (lower accuracy, better efficiency/cutting)}} \\
$\bullet$ & 44 &   4 & angle-y &       4 &      1 &  none     &       84.79 &  0.0570 &   1 \\
$\bullet$ & 40 &   4 & angle-x &       4 &      1 &  none     &       82.30 &  0.0535 &   1 \\
$\bullet$ & 06 &   1 & angle-x &       5 &      2 &  none-odd &       80.30 &  0.1231 &   4 \\
$\bullet$ & 69 &   6 & angle-y &       3 &      1 &  odd      &       72.19 &  0.0425 &   1 \\
$\bullet$ & 71 &   6 & angle-y &       4 &      2 &  none-odd &       70.24 &  0.0068 &   4 \\
\bottomrule
\end{tabular}%
}
\end{table}

\textbf{Key findings:}
Angle-y embedding with sparse CNOT (e.g., ``none-odd'') consistently outperforms other configurations; dense mode ``all'' clusters near random chance (${\sim}10\%$), suggesting barren plateaus. Depth-2 circuits offer the best accuracy cost tradeoff; 8 qubits achieve highest accuracy while intermediate sizes (4--6 qubits) often underperform smaller circuits.

\begin{figure}
  \centering
  \includegraphics[width=\linewidth]{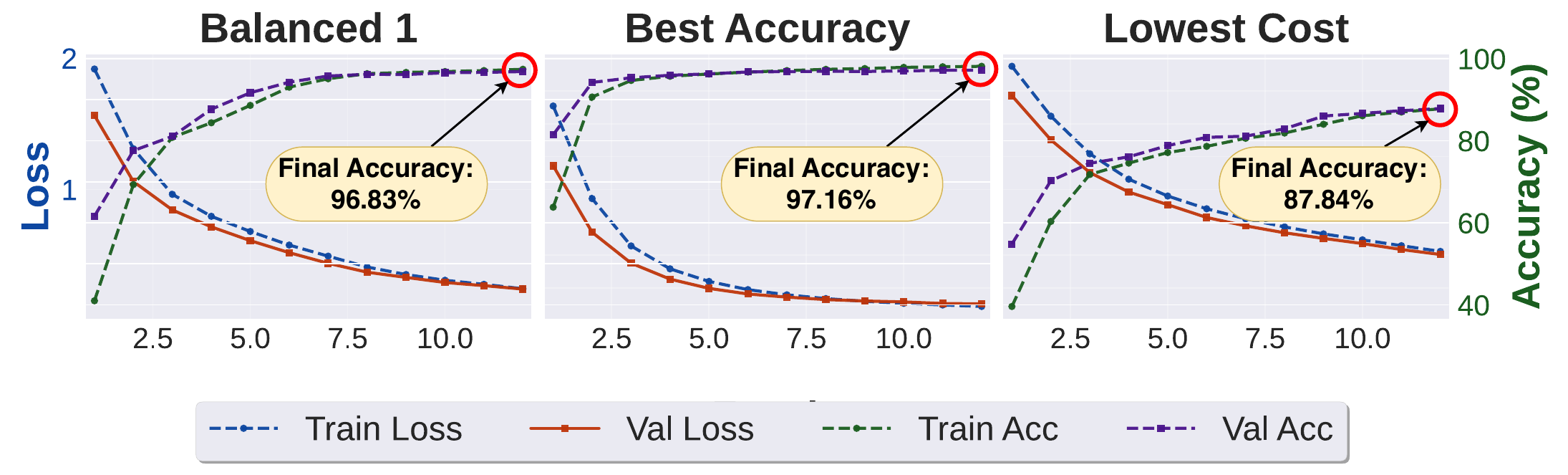}
  \caption{Training dynamics during 12 epoch retraining on full MNIST for three Pareto optimal architectures: \emph{Balanced} (96.83\% accuracy), \emph{Best Accuracy} (97.16\%), and \emph{Lowest Cost} (87.84\%). All configurations use angle-y embedding with CNOT mode ``none-odd''. Loss (left axis) and accuracy (right axis) are shown for both training and validation sets.}
  \label{fig:final_training}
\end{figure}

\textbf{Final retraining:}
Three representative Pareto optimal architectures were retrained for 12 epochs on the full 60{,}000-sample training set.
The \emph{Best Accuracy} model achieved \textbf{97.16\%} test accuracy, the \emph{Balanced} model reached \textbf{96.83\%}, and the \emph{Lowest Cost} model attained \textbf{87.84\%}. 
This demonstrates that the few epoch search effectively identifies high quality architectures that scale well with extended training.
Fig.~\ref{fig:final_training} shows the training dynamics for all three configurations, with loss steadily decreasing and accuracy converging after 12 epochs.

\begin{figure}[t!]
  \centering
  \includegraphics[width=\linewidth]{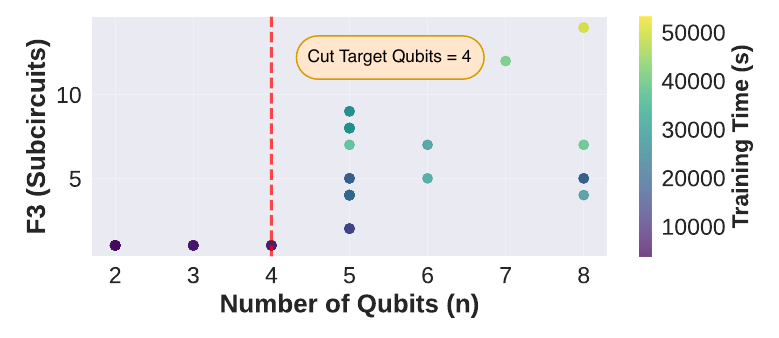}
  \caption{Number of subcircuits ($F_3$) versus qubit count for $Q_{\text{target}}{=}4$. Since cutting overhead scales as $\mathcal{O}(4^k)$ per cut, minimizing $F_3$ is critical.}
  \label{fig:mnist_f3_vs_n}
\end{figure}

\textbf{Cutting overhead analysis ($F_3$):}
With a fixed target qubit budget $Q_{\text{target}}{=}4$, circuits with $n \leq 4$ qubits and sparse entanglement require fewer subcircuits, while larger circuits (5--8 qubits) or those with denser entanglement patterns require more subcircuits (up to $F_3{=}14$ for complex 8 qubit, 4 layer circuits with mixed CNOT modes). Fig.~\ref{fig:mnist_f3_vs_n} visualizes this relationship.

\begin{figure}[t]
\centering
\small
\begin{minipage}[c]{0.64\linewidth}
\begin{minipage}{0.48\linewidth}
\centering
{\footnotesize\textbf{(Embed + Layer 1)}}
\end{minipage}
\hfill
\begin{minipage}{0.48\linewidth}
\centering
{\footnotesize\textbf{(Layer 2 + Meas)}}
\end{minipage}

\vspace{0.05cm}

\begin{minipage}{0.48\linewidth}
\centering
\tikz\node[draw=black, line width=1.5pt, fill=sc1bg, rounded corners=6pt, inner sep=3pt] {
\begin{minipage}[t][1.9cm]{0.92\linewidth}
\centering
{\footnotesize\textcolor{black}{SC1: $q_0$-$q_3$}}\\[0.15cm]
\raisebox{-0.6cm}[0pt][0pt]{%
\begin{adjustbox}{max width=\textwidth,max height=2.0cm,keepaspectratio=true}
\begin{quantikz}[row sep={0.7cm,between origins}, column sep=0.2cm]
\lstick{\large$q_0$} & \gate[style={fill=embedcolor,thick,rounded corners=2pt}]{\scriptsize R_Y} & \gate[style={fill=rxcolor,thick,rounded corners=2pt}]{\scriptsize R_X} & \gate[style={fill=rycolor,thick,rounded corners=2pt}]{\scriptsize R_Y} & \gate[style={fill=rzcolor,thick,rounded corners=2pt}]{\scriptsize R_Z} & \qw \rstick{\cutsym} \\
\lstick{\large$q_1$} & \gate[style={fill=embedcolor,thick,rounded corners=2pt}]{\scriptsize R_Y} & \gate[style={fill=rxcolor,thick,rounded corners=2pt}]{\scriptsize R_X} & \gate[style={fill=rycolor,thick,rounded corners=2pt}]{\scriptsize R_Y} & \gate[style={fill=rzcolor,thick,rounded corners=2pt}]{\scriptsize R_Z} & \qw \rstick{\cutsym} \\
\lstick{\large$q_2$} & \gate[style={fill=embedcolor,thick,rounded corners=2pt}]{\scriptsize R_Y} & \gate[style={fill=rxcolor,thick,rounded corners=2pt}]{\scriptsize R_X} & \gate[style={fill=rycolor,thick,rounded corners=2pt}]{\scriptsize R_Y} & \gate[style={fill=rzcolor,thick,rounded corners=2pt}]{\scriptsize R_Z} & \qw \rstick{\cutsym} \\
\lstick{\large$q_3$} & \gate[style={fill=embedcolor,thick,rounded corners=2pt}]{\scriptsize R_Y} & \gate[style={fill=rxcolor,thick,rounded corners=2pt}]{\scriptsize R_X} & \gate[style={fill=rycolor,thick,rounded corners=2pt}]{\scriptsize R_Y} & \gate[style={fill=rzcolor,thick,rounded corners=2pt}]{\scriptsize R_Z} & \qw \rstick{\cutsym} \\
\end{quantikz}
\end{adjustbox}%
}
\end{minipage}
};
\end{minipage}
\hfill
\begin{minipage}{0.48\linewidth}
\centering
\tikz\node[draw=black, line width=1.5pt, fill=sc3bg, rounded corners=6pt, inner sep=3pt] {
\begin{minipage}[t][1.9cm]{0.92\linewidth}
\centering
{\footnotesize\textcolor{black}{SC3: $q_0$-$q_3$}}\\[0.15cm]
\raisebox{-0.6cm}[0pt][0pt]{%
\begin{adjustbox}{max width=\textwidth,max height=2.0cm,keepaspectratio=true}
\begin{quantikz}[row sep={0.7cm,between origins}, column sep=0.2cm]
\lstick{\cutsym\large$q_0$} & \gate[style={fill=rxcolor,thick,rounded corners=2pt}]{\scriptsize R_X} & \gate[style={fill=rycolor,thick,rounded corners=2pt}]{\scriptsize R_Y} & \gate[style={fill=rzcolor,thick,rounded corners=2pt}]{\scriptsize R_Z} & \meter[style={fill=measurecolor,thick,rounded corners=2pt}]{} & \cw \\
\lstick{\cutsym\large$q_1$} & \gate[style={fill=rxcolor,thick,rounded corners=2pt}]{\scriptsize R_X} & \gate[style={fill=rycolor,thick,rounded corners=2pt}]{\scriptsize R_Y} & \gate[style={fill=rzcolor,thick,rounded corners=2pt}]{\scriptsize R_Z} & \qw \rstick{\cutsym} \\
\lstick{\cutsym\large$q_2$} & \gate[style={fill=rxcolor,thick,rounded corners=2pt}]{\scriptsize R_X} & \gate[style={fill=rycolor,thick,rounded corners=2pt}]{\scriptsize R_Y} & \gate[style={fill=rzcolor,thick,rounded corners=2pt}]{\scriptsize R_Z} & \meter[style={fill=measurecolor,thick,rounded corners=2pt}]{} & \cw \\
\lstick{\cutsym\large$q_3$} & \gate[style={fill=rxcolor,thick,rounded corners=2pt}]{\scriptsize R_X} & \gate[style={fill=rycolor,thick,rounded corners=2pt}]{\scriptsize R_Y} & \gate[style={fill=rzcolor,thick,rounded corners=2pt}]{\scriptsize R_Z} & \qw \rstick{\cutsym} \\
\end{quantikz}
\end{adjustbox}%
}
\end{minipage}
};
\end{minipage}

\vspace{0.08cm}

\begin{minipage}{0.48\linewidth}
\centering
\tikz\node[draw=black, line width=1.5pt, fill=sc2bg, rounded corners=6pt, inner sep=3pt] {
\begin{minipage}[t][1.9cm]{0.92\linewidth}
\centering
{\footnotesize\textcolor{black}{SC2: $q_4$-$q_7$}}\\[0.15cm]
\raisebox{-0.6cm}[0pt][0pt]{%
\begin{adjustbox}{max width=\textwidth,max height=2.0cm,keepaspectratio=true}
\begin{quantikz}[row sep={0.7cm,between origins}, column sep=0.2cm]
\lstick{\large$q_4$} & \gate[style={fill=embedcolor,thick,rounded corners=2pt}]{\scriptsize R_Y} & \gate[style={fill=rxcolor,thick,rounded corners=2pt}]{\scriptsize R_X} & \gate[style={fill=rycolor,thick,rounded corners=2pt}]{\scriptsize R_Y} & \gate[style={fill=rzcolor,thick,rounded corners=2pt}]{\scriptsize R_Z} & \qw \rstick{\cutsym} \\
\lstick{\large$q_5$} & \gate[style={fill=embedcolor,thick,rounded corners=2pt}]{\scriptsize R_Y} & \gate[style={fill=rxcolor,thick,rounded corners=2pt}]{\scriptsize R_X} & \gate[style={fill=rycolor,thick,rounded corners=2pt}]{\scriptsize R_Y} & \gate[style={fill=rzcolor,thick,rounded corners=2pt}]{\scriptsize R_Z} & \qw \rstick{\cutsym} \\
\lstick{\large$q_6$} & \gate[style={fill=embedcolor,thick,rounded corners=2pt}]{\scriptsize R_Y} & \gate[style={fill=rxcolor,thick,rounded corners=2pt}]{\scriptsize R_X} & \gate[style={fill=rycolor,thick,rounded corners=2pt}]{\scriptsize R_Y} & \gate[style={fill=rzcolor,thick,rounded corners=2pt}]{\scriptsize R_Z} & \qw \rstick{\cutsym} \\
\lstick{\large$q_7$} & \gate[style={fill=embedcolor,thick,rounded corners=2pt}]{\scriptsize R_Y} & \gate[style={fill=rxcolor,thick,rounded corners=2pt}]{\scriptsize R_X} & \gate[style={fill=rycolor,thick,rounded corners=2pt}]{\scriptsize R_Y} & \gate[style={fill=rzcolor,thick,rounded corners=2pt}]{\scriptsize R_Z} & \qw \rstick{\cutsym} \\
\end{quantikz}
\end{adjustbox}%
}
\end{minipage}
};
\end{minipage}
\hfill
\begin{minipage}{0.48\linewidth}
\centering
\tikz\node[draw=black, line width=1.5pt, fill=sc4bg, rounded corners=6pt, inner sep=3pt] {
\begin{minipage}[t][1.9cm]{0.92\linewidth}
\centering
{\footnotesize\textcolor{black}{SC4: $q_4$-$q_7$}}\\[0.15cm]
\raisebox{-0.6cm}[0pt][0pt]{%
\begin{adjustbox}{max width=\textwidth,max height=2.0cm,keepaspectratio=true}
\begin{quantikz}[row sep={0.7cm,between origins}, column sep=0.2cm]
\lstick{\cutsym\large$q_4$} & \gate[style={fill=rxcolor,thick,rounded corners=2pt}]{\scriptsize R_X} & \gate[style={fill=rycolor,thick,rounded corners=2pt}]{\scriptsize R_Y} & \gate[style={fill=rzcolor,thick,rounded corners=2pt}]{\scriptsize R_Z} & \meter[style={fill=measurecolor,thick,rounded corners=2pt}]{} & \cw \\
\lstick{\cutsym\large$q_5$} & \gate[style={fill=rxcolor,thick,rounded corners=2pt}]{\scriptsize R_X} & \gate[style={fill=rycolor,thick,rounded corners=2pt}]{\scriptsize R_Y} & \gate[style={fill=rzcolor,thick,rounded corners=2pt}]{\scriptsize R_Z} & \qw \rstick{\cutsym} \\
\lstick{\cutsym\large$q_6$} & \gate[style={fill=rxcolor,thick,rounded corners=2pt}]{\scriptsize R_X} & \gate[style={fill=rycolor,thick,rounded corners=2pt}]{\scriptsize R_Y} & \gate[style={fill=rzcolor,thick,rounded corners=2pt}]{\scriptsize R_Z} & \meter[style={fill=measurecolor,thick,rounded corners=2pt}]{} & \cw \\
\lstick{\cutsym\large$q_7$} & \gate[style={fill=rxcolor,thick,rounded corners=2pt}]{\scriptsize R_X} & \gate[style={fill=rycolor,thick,rounded corners=2pt}]{\scriptsize R_Y} & \gate[style={fill=rzcolor,thick,rounded corners=2pt}]{\scriptsize R_Z} & \qw \rstick{\cutsym} \\
\end{quantikz}
\end{adjustbox}%
}
\end{minipage}
};
\end{minipage}
\end{minipage}
\hfill
\begin{minipage}[c]{0.33\linewidth}
\centering
{\footnotesize\textbf{(CNOTs + Meas)}}

\vspace{0.05cm}

\tikz\node[draw=black, line width=1.5pt, fill=sc5bg, rounded corners=6pt, inner sep=3pt] {
\begin{minipage}[c][4.08cm]{0.92\linewidth}
\centering
{\footnotesize\textcolor{black}{SC5: $q_1,q_3,q_5,q_7$}}\\[0.2cm]
\begin{adjustbox}{max width=\linewidth,max height=2.8cm,keepaspectratio=true}
\begin{quantikz}[row sep={0.6cm,between origins}, column sep=0.35cm]
\lstick{\cutsym\large$q_1$} & \ctrl{2} & \qw & \targ{} & \qw & \meter[style={fill=measurecolor,thick,rounded corners=2pt}]{} & \cw \\
\lstick{\cutsym\large$q_3$} & \qw & \ctrl{2} & \qw & \targ{} & \meter[style={fill=measurecolor,thick,rounded corners=2pt}]{} & \cw \\
\lstick{\cutsym\large$q_5$} & \targ{} & \qw & \ctrl{-2} & \qw & \meter[style={fill=measurecolor,thick,rounded corners=2pt}]{} & \cw \\
\lstick{\cutsym\large$q_7$} & \qw & \targ{} & \qw & \ctrl{-2} & \meter[style={fill=measurecolor,thick,rounded corners=2pt}]{} & \cw \\
\end{quantikz}
\end{adjustbox}
\end{minipage}
};
\end{minipage}

\vspace{0.1cm}
\fbox{\footnotesize
\tikz\fill[embedcolor,rounded corners=1pt] (0,0) rectangle (0.25,0.15); Embed~
\tikz\fill[rxcolor,rounded corners=1pt] (0,0) rectangle (0.25,0.15); $R_X$~
\tikz\fill[rycolor,rounded corners=1pt] (0,0) rectangle (0.25,0.15); $R_Y$~
\tikz\fill[rzcolor,rounded corners=1pt] (0,0) rectangle (0.25,0.15); $R_Z$~
\tikz\fill[measurecolor,rounded corners=1pt] (0,0) rectangle (0.25,0.15); Meas~
\cutsym~Cut
}

\caption{Circuit cutting for the best 8q-2d angle-y architecture ($F_3{=}5$): 5 subcircuits for a 4 qubit device. SC1-2: embedding and Layer~1; SC3-4: Layer~2; SC5: cross group CNOTs and measurements.}
\label{fig:circuit_cutting}
\end{figure}
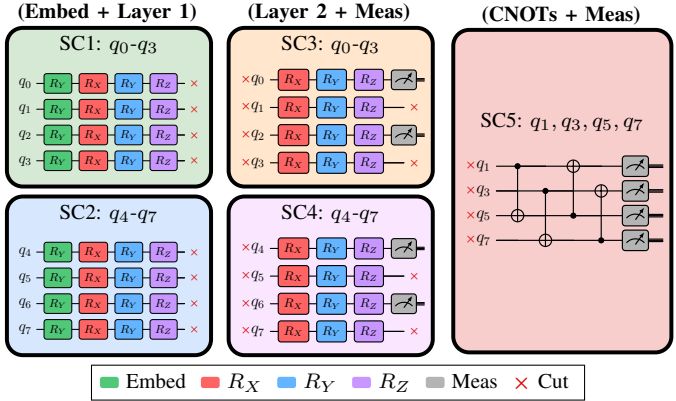

Fig.~\ref{fig:circuit_cutting} illustrates how the best performing 8 qubit circuit is partitioned into 5 subcircuits for deployment on a 4 qubit device: SC1-2 handle embedding and Layer~1, SC3-4 execute Layer~2, and SC5 performs cross group CNOTs followed by measurements.

\textbf{Transpilation analysis:}
To assess hardware deployability, we transpile top-performing architectures to IBM Quantum backends using Qiskit's fake provider snapshots (Sec.~\ref{Sec_Transpilation}). Table~\ref{tab:transpilation} reports pre- and post-transpilation metrics for representative architectures. 

\begin{table}[t!]
\caption{Transpilation analysis for Pareto optimal architectures. Circuits transpiled to \texttt{fake\_melbourne} (15q) with optimization level 3. Sparse CNOT modes yield minimal routing overhead.}
\label{tab:transpilation}
  \centering
\small
\setlength{\tabcolsep}{2.5pt}
\begin{tabular}{lccccccl}
\toprule
Config & \multicolumn{2}{c}{Depth} & \multicolumn{2}{c}{CNOTs} & \multicolumn{2}{c}{Gates} & CNOT \\
\cmidrule(lr){2-3} \cmidrule(lr){4-5} \cmidrule(lr){6-7}
 & Pre & Post & Pre & Post & Pre & Post & Mode \\
\midrule
8q-2d (Best Acc.) & 10 & 14 & 4 & 4 & 68 & 68 & none-odd \\
5q-2d (Balanced)  & 9 & 12 & 2 & 2 & 42 & 42 & none-odd \\
4q-2d (Low Cost)  & 9 & 12 & 2 & 2 & 34 & 36 & none-odd \\
\bottomrule
\end{tabular}
\end{table}

For the best 8-qubit, 2-layer architecture with CNOT mode ``none-odd'', the sparse entangling pattern (only 4 CNOTs) results in manageable transpilation overhead, circuit depth increases from 10 to 14 due to routing, while CNOT count remains unchanged at 4. Fig.~\ref{fig:transpiled_circuit} shows the transpiled circuit mapped to \texttt{fake\_melbourne}, with the transpiler decomposing single qubit rotations into the native gate set ($R_z$ and $\sqrt{X}$) while preserving the CNOT count.

\begin{figure}
  \centering
  \begin{adjustbox}{max width=\linewidth,scale=1}
    \begin{quantikz}[row sep={0.9cm,between origins}, column sep=0.32cm]
      \lstick{\large$q_5 \mapsto 0$}
        & \gate[style=rzgate]{\substack{R_Z \\ \scriptstyle -2.51}}
        & \gate[style=sqrtxgate]{\sqrt{X}}
        & \gate[style=rzgate]{\substack{R_Z \\ \scriptstyle -1.93}}
        & \gate[style=sqrtxgate]{\sqrt{X}}
        & \gate[style=rzgate]{\substack{R_Z \\ \scriptstyle -\pi/2}}
        & \targ{}
        & \gate[style=rzgate]{\substack{R_Z \\ \scriptstyle \pi/2}}
        & \qw
        & \targ{}
        & \gate[style=sqrtxgate]{\sqrt{X}}
        & \gate[style=rzgate]{\substack{R_Z \\ \scriptstyle 1.42}}
        & \gate[style=sqrtxgate]{\sqrt{X}}
        & \gate[style=rzgate]{\substack{R_Z \\ \scriptstyle -\pi/2}}
        & \qw
        & \meter[style=metergate]{}
      \\
      \lstick{\large$q_1 \mapsto 1$}
        & \gate[style=rzgate]{\substack{R_Z \\ \scriptstyle 0.176}}
        & \gate[style=sqrtxgate]{\sqrt{X}}
        & \gate[style=rzgate]{\substack{R_Z \\ \scriptstyle 0.324}}
        & \gate[style=sqrtxgate]{\sqrt{X}}
        & \qw
        & \ctrl{-1}
        & \gate[style=rzgate]{\substack{R_Z \\ \scriptstyle -\pi/2}}
        & \gate[style=sqrtxgate]{\sqrt{X}}
        & \ctrl{-1}
        & \gate[style=sqrtxgate]{\sqrt{X}}
        & \gate[style=rzgate]{\substack{R_Z \\ \scriptstyle -0.427}}
        & \qw & \qw & \qw
        & \meter[style=metergate]{}
      \\
      \lstick{\large$q_2 \mapsto 2$}
        & \gate[style=rzgate]{\substack{R_Z \\ \scriptstyle 0.629}}
        & \gate[style=sqrtxgate]{\sqrt{X}}
        & \gate[style=rzgate]{\substack{R_Z \\ \scriptstyle -2.78}}
        & \gate[style=sqrtxgate]{\sqrt{X}}
        & \gate[style=rzgate]{\substack{R_Z \\ \scriptstyle 2.61}}
        & \qw & \qw & \qw & \qw & \qw & \qw & \qw & \qw & \qw
        & \meter[style=metergate]{}
      \\
      \lstick{\large$q_4 \mapsto 4$}
        & \gate[style=rzgate]{\substack{R_Z \\ \scriptstyle 0.629}}
        & \gate[style=sqrtxgate]{\sqrt{X}}
        & \gate[style=rzgate]{\substack{R_Z \\ \scriptstyle -2.78}}
        & \gate[style=sqrtxgate]{\sqrt{X}}
        & \gate[style=rzgate]{\substack{R_Z \\ \scriptstyle 2.61}}
        & \qw & \qw & \qw & \qw & \qw & \qw & \qw & \qw & \qw
        & \meter[style=metergate]{}
      \\
      \lstick{\large$q_0 \mapsto 8$}
        & \gate[style=rzgate]{\substack{R_Z \\ \scriptstyle 0.629}}
        & \gate[style=sqrtxgate]{\sqrt{X}}
        & \gate[style=rzgate]{\substack{R_Z \\ \scriptstyle -2.78}}
        & \gate[style=sqrtxgate]{\sqrt{X}}
        & \gate[style=rzgate]{\substack{R_Z \\ \scriptstyle 2.61}}
        & \qw & \qw & \qw & \qw & \qw & \qw & \qw & \qw & \qw
        & \meter[style=metergate]{}
      \\
      \lstick{\large$q_7 \mapsto 10$}
        & \gate[style=rzgate]{\substack{R_Z \\ \scriptstyle 0.629}}
        & \gate[style=sqrtxgate]{\sqrt{X}}
        & \gate[style=rzgate]{\substack{R_Z \\ \scriptstyle -1.21}}
        & \gate[style=sqrtxgate]{\sqrt{X}}
        & \qw
        & \ctrl{1}
        & \gate[style=rzgate]{\substack{R_Z \\ \scriptstyle -\pi/2}}
        & \gate[style=sqrtxgate]{\sqrt{X}}
        & \ctrl{1}
        & \gate[style=sqrtxgate]{\sqrt{X}}
        & \gate[style=rzgate]{\substack{R_Z \\ \scriptstyle -1.72}}
        & \gate[style=sqrtxgate]{\sqrt{X}}
        & \gate[style=rzgate]{\substack{R_Z \\ \scriptstyle -\pi/2}}
        & \qw
        & \meter[style=metergate]{}
      \\
      \lstick{\large$q_3 \mapsto 11$}
        & \gate[style=rzgate]{\substack{R_Z \\ \scriptstyle 0.176}}
        & \gate[style=sqrtxgate]{\sqrt{X}}
        & \gate[style=rzgate]{\substack{R_Z \\ \scriptstyle -2.82}}
        & \gate[style=sqrtxgate]{\sqrt{X}}
        & \gate[style=rzgate]{\substack{R_Z \\ \scriptstyle \pi/2}}
        & \targ{}
        & \gate[style=rzgate]{\substack{R_Z \\ \scriptstyle \pi/2}}
        & \qw
        & \targ{}
        & \gate[style=sqrtxgate]{\sqrt{X}}
        & \gate[style=rzgate]{\substack{R_Z \\ \scriptstyle 2.71}}
        & \qw & \qw & \qw
        & \meter[style=metergate]{}
      \\
      \lstick{\large$q_6 \mapsto 12$}
        & \gate[style=rzgate]{\substack{R_Z \\ \scriptstyle 0.629}}
        & \gate[style=sqrtxgate]{\sqrt{X}}
        & \gate[style=rzgate]{\substack{R_Z \\ \scriptstyle -2.78}}
        & \gate[style=sqrtxgate]{\sqrt{X}}
        & \gate[style=rzgate]{\substack{R_Z \\ \scriptstyle 2.61}}
        & \qw & \qw & \qw & \qw & \qw & \qw & \qw & \qw & \qw
        & \meter[style=metergate]{}
      \\
    \end{quantikz}
  \end{adjustbox}
  \caption{Transpiled 8 qubit angle-y circuit (CNOT mode ``none-odd'') on \texttt{fake\_melbourne} with global phase $\pi$. The transpiler decomposes rotations into the native basis $\{R_z, \sqrt{X}\}$. The mapping $q_i \mapsto j$ shows logical to physical qubit assignment.}
  \label{fig:transpiled_circuit}
\end{figure}
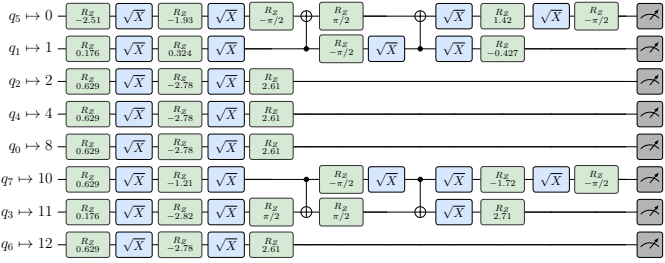

\subsection{Results on Fashion-MNIST}
\label{Sec_Results_FashionMNIST}

To validate QNAS on a more challenging benchmark, we apply the same search protocol to Fashion-MNIST. This dataset shares the same structure as MNIST (60{,}000 training / 10{,}000 test images, 10 classes) but features clothing items instead of digits, presenting greater intra class variation and inter class similarity.

\textbf{Search setup:}
Using the same search space (Table~\ref{tab:search_space}), NSGA-II evaluated \textbf{66} HQNN candidates over six generations with 2 epoch training per candidate. Table~\ref{tab:fmnist_summary} summarizes the search phase.

\begin{table}[t!]
\caption{Fashion-MNIST search phase summary.}
\label{tab:fmnist_summary}
\centering
\small
\setlength{\tabcolsep}{3pt}
\resizebox{\linewidth}{!}{%
\begin{tabular}{llll}
\toprule
                        Metric &               Value &                        Metric &               Value \\
\midrule
         Total evaluations &                   66 & Best search val accuracy (\%) &               83.28 \\
                Generations &                    6 & Best final test accuracy (\%) &               87.38 \\
         Training epochs/eval &                    2 &   Mean $\pm$ std accuracy (\%) &       52.91 $\pm$ 26.47 \\
         $Q_{\text{target}}$ (qubits) &                    4 &      Mean $F_2$ (s/sample) &               0.066 \\
      Total search time (h) &               65.51 & & \\
\bottomrule
\end{tabular}%
}
\end{table}

\textbf{Top configurations:}
Table~\ref{tab:fmnist_results} lists the top configurations and Pareto optimal solutions. The best search phase model achieved \textbf{83.28\%} validation accuracy with configuration 5q-2d angle-y, CNOT mode none-none.

\begin{table}
\caption{Fashion-MNIST search results: top configurations by accuracy and Pareto optimal solutions. Angle-y embedding dominates across all top performers.}
\label{tab:fmnist_results}
\centering
\small
\setlength{\tabcolsep}{2.5pt}
\resizebox{\linewidth}{!}{%
\begin{tabular}{clrrrllrrr}
\toprule
 & EvalID & Gen &  Embed &  Qubits &  Depth &  CNOT Mode &  ValAcc(\%) &  $F_2$ (s/samp) &  $F_3$ \\
\midrule
\multicolumn{10}{l}{\textit{Top accuracy configurations}} \\
$\bullet$ & 31 &   3 & angle-y &       5 &      2 &  none-none &       83.28 &  0.0694 &   3 \\
          & 63 &   6 & angle-x &       5 &      1 &  none      &       81.88 &  0.0805 &   2 \\
$\bullet$ & 47 &   5 & angle-y &       8 &      4 &  none-none-odd-odd &       80.15 &  0.0258 &  12 \\
          & 67 &   6 & angle-y &       5 &      2 &  none-none &       80.13 &  0.1158 &   3 \\
$\bullet$ & 22 &   2 & angle-y &       6 &      1 &  even      &       79.74 &  0.0075 &   2 \\
          & 54 &   5 & angle-y &       5 &      1 &  odd       &       79.81 &  0.0842 &   2 \\
\midrule
\multicolumn{10}{l}{\textit{Additional Pareto optimal solutions}} \\
$\bullet$ & 36 &   3 & angle-y &       5 &      1 &  odd       &       64.06 &  0.0065 &   2 \\
$\bullet$ & 05 &   1 & angle-y &       5 &      1 &  odd       &       58.01 &  0.0064 &   2 \\
\bottomrule
\end{tabular}%
}
\end{table}

\textbf{Final retraining:}
Selected Pareto optimal architectures were retrained for 12 epochs on the full training set. The \emph{Best Accuracy} model (5q-2d, angle-y, CNOT mode none-none) achieved \textbf{87.38\%} test accuracy, while the \emph{Balanced} model (8q-4d) reached \textbf{87.24\%}. Lower cost configurations achieved 75--77\% accuracy.

\textbf{Consistency with MNIST findings:}
Angle-y and sparse CNOT again dominate; lower accuracy (87.38\% vs.\ 97.16\%) reflects the harder dataset.

\subsection{Results on Iris Dataset}
\label{Sec_Results_Iris}

To demonstrate QNAS's applicability beyond grayscale image classification, we evaluate on the Iris dataset (3 class classification with 4 dimensional features). NSGA-II evaluated \textbf{120} HQNN candidates over ten generations with 5 epoch training per candidate. Table~\ref{tab:iris_summary} summarizes the search phase.

\begin{table}
\caption{Iris search space configuration.}
\label{tab:iris_search_space}
\centering
\small
\resizebox{\linewidth}{!}{%
\begin{tabular}{ll}
\toprule
\textbf{Parameter} & \textbf{Range} \\
\midrule
Embedding type & angle-$X$, angle-$Y$, angle-$Z$, amplitude \\
Qubits $n$ & $[2, 8]$ \\
Depth $L$ & $[1, 5]$ \\
Entanglement range $r$ (per layer) & $[1, n{-}1]$ \\
CNOT mode (per layer) & all, odd, even, none \\
Learning rate $\eta$ & $[10^{-3}, 5{\times}10^{-3}]$ \\
$Q_{\text{target}}$ (qubits) & 2 \\
\bottomrule
\end{tabular}
}
\end{table}

\begin{table}[t!]
\caption{Iris search phase summary.}
\label{tab:iris_summary}
\centering
\small
\setlength{\tabcolsep}{3pt}
\resizebox{\linewidth}{!}{%
\begin{tabular}{llll}
\toprule
                        Metric &               Value &                        Metric &               Value \\
\midrule
         Total evaluations &                  120 & Best search val accuracy (\%) &              100.00 \\
                Generations &                   10 & Best final val accuracy (\%) &              100.00 \\
         Training epochs/eval &                    5 &   Mean $\pm$ std accuracy (\%) &       66.67 $\pm$ 28.19 \\
         $Q_{\text{target}}$ (qubits) &                    2 &      Mean $F_2$ (s/sample) &               0.121 \\
      Total search time (h) &                0.60 & & \\
\bottomrule
\end{tabular}%
}
\end{table}

\textbf{Top configurations:}
The best search phase model achieved \textbf{100\%} validation accuracy with configuration 4q-2d amplitude, CNOT mode none-odd, $F_2{=}0.0070$\,s/sample, $F_3{=}6$. Table~\ref{tab:iris_results} lists the top configurations by accuracy and selected Pareto optimal solutions.

\begin{table}[t!]
\vspace*{3pt}
\caption{Iris search results: top configurations by accuracy and selected Pareto optimal solutions. Amplitude embedding dominates; best accuracy is 4q-2d with none-odd.}
\label{tab:iris_results}
\centering
\small
\setlength{\tabcolsep}{2.5pt}
\resizebox{\linewidth}{!}{%
\begin{tabular}{clrrrllrrr}
\toprule
 & EvalID & Gen &  Embed &  Qubits &  Depth &  CNOT Mode &  ValAcc(\%) &  $F_2$ (s/samp) &  $F_3$ \\
\midrule
\multicolumn{10}{l}{\textit{Top accuracy configurations}} \\
$\bullet$ & 62 &   5 & amplitude &       4 &      2 &  none-odd &      100.00 &  0.0070 &   6 \\
$\bullet$ & 104 &   8 & amplitude &       2 &      2 &  odd-all &      100.00 &  0.0634 &   1 \\
          & 48 &   3 & amplitude &       4 &      2 &  none-even &      100.00 &  0.0395 &   6 \\
          & 71 &   5 & amplitude &       4 &      2 &  none-even &      100.00 &  0.0373 &   6 \\
$\bullet$ & 74 &   6 & amplitude &       4 &      1 &  odd &      100.00 &  0.0256 &   4 \\
          & 84 &   6 & amplitude &       4 &      2 &  none-odd &       96.67 &  0.0070 &   6 \\
          & 78 &   6 & amplitude &       4 &      2 &  none-even &       96.67 &  0.0074 &   6 \\
          & 95 &   7 & amplitude &       4 &      1 &  odd &      100.00 &  0.0148 &   2 \\
\midrule
\multicolumn{10}{l}{\textit{Additional Pareto optimal (low cost / low $F_3$)}} \\
$\bullet$ & 81 &   6 & angle-x &       2 &      1 &  none &       96.67 &  0.0528 &   1 \\
$\bullet$ & 114 &   9 & amplitude &       2 &      1 &  none &       90.00 &  0.0355 &   1 \\
\bottomrule
\end{tabular}%
}
\end{table}

\textbf{Key findings and final retraining:}
In contrast to MNIST and Fashion-MNIST, where angle-y embedding dominates, amplitude embedding performs best on Iris. This is due to the small scale of the dataset (150 samples, 4 features) and the use of a compact classical head: one pre- and one post-classical layer with hidden dimension 16, so the number of trainable parameters does not explode despite the higher dimensionality of amplitude encoding. Sparse CNOT modes (e.g., none-odd, none-even) dominate. The number of subcircuits ($F_3$) varies when partitioned for the 2 qubit target ($Q_{\text{target}}{=}2$). Selected Pareto optimal architectures were retrained for 20 epochs on the full training set. The \emph{Best Accuracy} model (4q-2d, amplitude, none-odd) achieved \textbf{100\%} validation accuracy after retraining, demonstrating QNAS's ability to discover cutting-efficient architectures across different data modalities.

\section{Conclusion}
\label{Sec_Conclusion}

We introduced QNAS, a neural architecture search framework for hybrid quantum-classical neural networks that jointly optimizes accuracy, runtime cost, and circuit cutting overhead. Our key findings across MNIST, Fashion-MNIST, and Iris include: (1) sparse CNOT modes consistently outperform dense entanglement patterns across all datasets, with embedding type (angle-y for images, amplitude for Iris) and optimal CNOT patterns varying by domain; (2) few-epoch training reliably ranks candidate architectures with $>$0.9 correlation to final performance, enabling efficient evolutionary search; and (3) hardware-friendly architectures discovered by QNAS require minimal transpilation overhead and cutting overhead. The best architectures achieved 97.16\% test accuracy on MNIST (\mbox{8-qubit}, \mbox{2-layer}), 87.38\% on Fashion-MNIST (\mbox{5-qubit}, \mbox{2-layer}), and 100\% validation accuracy on Iris (\mbox{4-qubit}, \mbox{2-layer}), demonstrating QNAS's effectiveness across different data modalities.

\textbf{Limitations.} Our evaluation focuses on image classification and tabular data; generalization to additional domains remains to be validated. While our transpilation to snapshot based backends incorporates realistic noise models, validation on actual quantum hardware remains as future work.

\textbf{Future work.} We plan to: (i) extend QNAS to additional domains beyond image classification and tabular data; (ii) execute on real quantum devices to validate performance under actual hardware noise; (iii) explore alternative search strategies, such as differentiable NAS; and (iv) conduct multi budget analysis by varying $Q_{\text{target}}$ to expose tradeoffs under different hardware constraints.

\section*{Acknowledgment}
This work was supported in part by the NYUAD Center for Quantum and Topological Systems (CQTS), funded by Tamkeen under the NYUAD Research Institute grant CG008.




\bibliographystyle{IEEEtran}
\bibliography{main}

\end{document}